\pdfoutput=1

\documentclass[11pt]{article}

\usepackage[preprint]{acl}

\usepackage{times}
\usepackage{latexsym}
\usepackage{amsmath}
\usepackage{amsthm,amssymb,multirow}
\usepackage{booktabs}
\usepackage{placeins}
\usepackage{algpseudocode}
\usepackage{algorithm}
\usepackage{listings,tikz}
\usepackage[most]{tcolorbox}
\usepackage{mathbbol}
\usepackage{calc}
\usepackage{subcaption}
\usepackage{comment,xcolor,soul}
\usepackage[frozencache,cachedir=minted-cache]{minted}
\tcbuselibrary{minted}
\sethlcolor{lightblue}
\usepackage{verbatim}
\usepackage{hyperref}
\usepackage{kotex}
\usepackage{fontawesome}

\newtcblisting{pythoncode}{
  listing engine=minted,
  breakable,
  listing only,
  nobeforeafter,
  minted style=colorful,
  minted language=python,
  enhanced,
  minted options = {
        breaklines=true, 
        fontsize=\small
      },
}

\newtcblisting{javacode}{
  listing engine=minted,
  breakable,
  listing only,
  minted style=colorful,
  minted language=java,
  enhanced,
  minted options = {
        breaklines=true, 
        fontsize=\small
      },
}

\usepackage[T1]{fontenc}
\usepackage[utf8]{inputenc}

\usepackage{microtype}

\usepackage{inconsolata}

\usepackage{graphicx}

\DeclareMathOperator*{\argmax}{arg\,max}

\title{TCProF: Time-Complexity Prediction SSL Framework}

\author{
 \textbf{Joonghyuk Hahn},
 \textbf{Hyeseon Ahn},
 \textbf{Jungin Kim},
 \textbf{Soohan Lim},
 \textbf{Yo-Sub Han}\thanks{Corresponding author.},
 \\
 Department of Computer Science, Yonsei University, Seoul, Republic of Korea,
\\
   \texttt{\{%
   \href{mailto:greghahn@yonsei.ac.kr}{greghahn},%
   \href{mailto:hsan@yonsei.ac.kr}{hsan},%
   \href{mailto:jungin.kim@yonsei.ac.kr}{jungin.kim},%
   \href{mailto:aness1219@yonsei.ac.kr}{aness1219},%
   \href{mailto:emmous@yonsei.ac.kr}{emmous}%
   \}@yonsei.ac.kr}
}

\begin{document}
\maketitle
\begin{abstract}
Time complexity is a theoretic measure to determine
the amount of time the algorithm needs for its execution.
In reality, developers write algorithms into code snippets within limited resources,
making the calculation of a code's time complexity a fundamental task.
However, determining the precise time complexity of a code is theoretically undecidable.
In response,
recent advancements have leaned toward deploying datasets for code time complexity prediction
and initiating preliminary experiments for this challenge.
We investigate the challenge in low-resource scenarios where
only a few labeled instances are given for training.
Remarkably, we are the first to introduce
\texttt{TCProF}: a \textbf{T}ime-\textbf{C}omplexity \textbf{Pr}edicti\textbf{o}n SSL \textbf{F}ramework
as an effective solution for code time complexity prediction in low-resource settings.
\texttt{TCProF} significantly boosts performance by integrating our
augmentation, symbolic modules, and a co-training mechanism,
achieving a more than 60\% improvement over self-training approaches.
We further provide an extensive comparative analysis between \texttt{TCProF},
ChatGPT, and Gemini-Pro, offering a detailed evaluation of our approach. \href{https://github.com/peer0/few-shot-tc}{\faGithub}
\end{abstract}

\section{Introduction}
The task of predicting time complexity for code snippets represents a significant challenge
in programming efficiency analysis.
Time complexity is a crucial benchmark for evaluating algorithm performance across diverse computational domains.
However, accurately computing the time complexity of a code snippet remains theoretically undecidable~\cite{Asperti08},
presenting a substantial obstacle.
This issue is particularly crucial in environments such as
educational settings, programming competitions, and automated code reviews,
where an accurate evaluation of numerous solutions is essential.

In the meantime, the advent of deep learning methodologies presents a promising avenue to address this challenge.
\citet{SikkaSKUSZ20} introduced \textrm{CorCoD} dataset, specifically designed for code time complexity prediction.
The dataset consists of code snippets labeled with their time complexity classes.
They also provide initial experiments for the code time complexity prediction
using both conventional algorithms and basic neural models.
Despite these advances~\cite{BaikJHKHK24,MoudgalyaRCL23}, the efficiency of such models heavily relies on the availability of
extensively annotated datasets.
Unfortunately, datasets in this domain are currently scarce due to the problem of data scarcity,
as time complexity annotations require professional knowledge.
While there are further approaches~\cite{BaikJHKHK24,MoudgalyaRCL23} with considerable potential,
their effectiveness is contingent upon the size of annotated datasets.

Addressing the shortage of labeled data, we introduce several innovations.
We develop data augmentation techniques specifically tailored to identify
key factors that influence the time complexity of the code snippets.
We also incorporate a co-training mechanism that leverages both original and augmented data effectively.
Additionally, we construct a symbolic module that enhances the accuracy of pseudo-labels
compared to the pseudo-labels generated by a model-based approach alone.
Collectively, these components form the backbone of \texttt{TCProF},
our semi-supervised learning~(SSL) framework,
which is uniquely equipped to address code time complexity prediction in low-resource settings.

\begin{figure*}[ht]
    \centering
    \includegraphics[width=\textwidth]{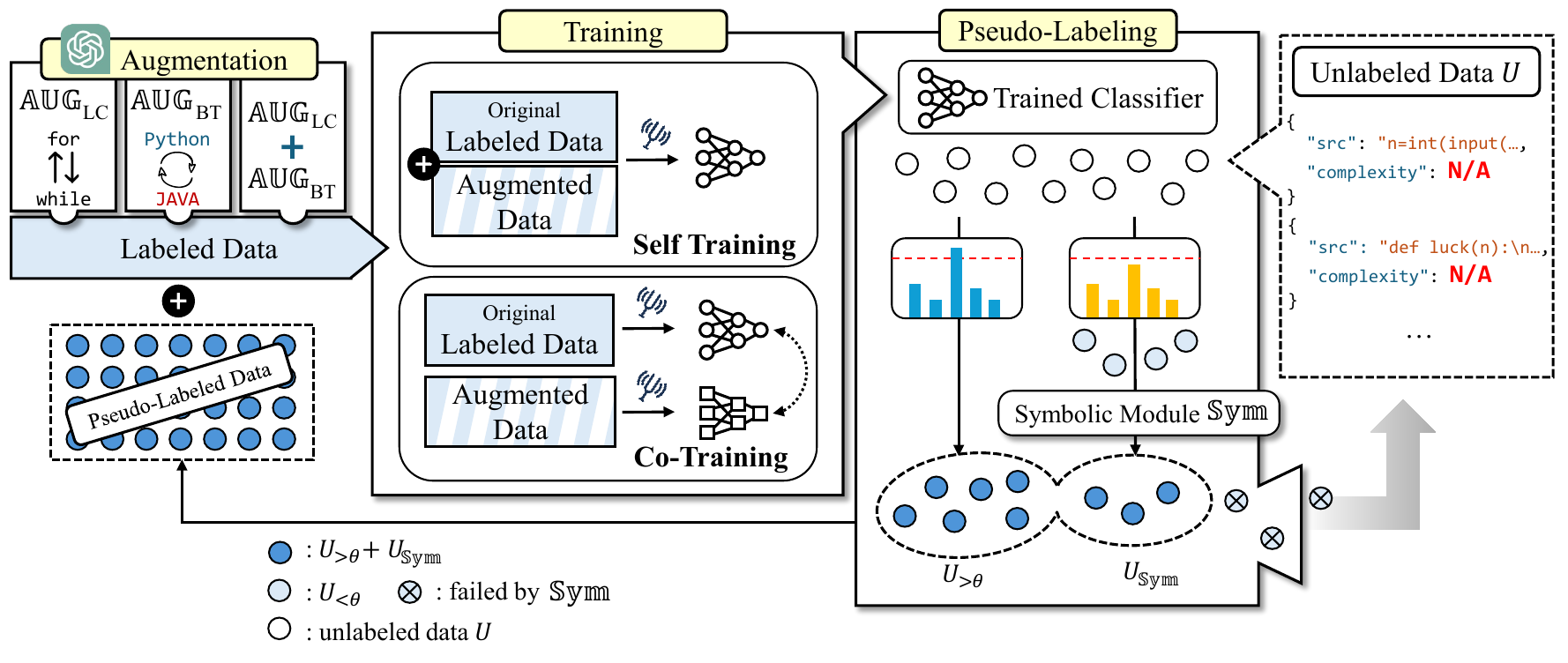}
    \caption{The overall framework of \texttt{TCProF}.}
    \label{fig:framework}
\end{figure*}

Operating under the assumption of limited labeled data and a vast number of unlabeled code snippets,
we empirically analyze \texttt{TCProF} using publicly available datasets, CorCoD and \textrm{CodeComplex}~\cite{BaikJHKHK24}.
CorCoD includes code snippets categorized into five different complexity classes, $O(1)$, $O(\log N)$, $O(N)$, $O(N\log N)$, and $O(N^2)$.
In contrast, CodeComplex consists of codes across seven different complexity classes, $O(1)$, $O(\log N)$, $O(N)$, $O(N\log N)$, $O(N^2)$, $O(N^3)$, and $O(2^N)$.
For these benchmark datasets,
our framework~\texttt{TCProF}
significantly enhances the performance over traditional self-training methods with
improvements exceeding 60\% in accuracy and F1-scores.
In the era of ChatGPT,
we further provide comparative studies against off-the-shelf large language models~(LLMs).
Meanwhile, given the early stage of research in this domain,
we provide an in-depth analysis of \texttt{TCProF} offering valuable insights into their practical utility.
Ultimately, our endeavor specifically targets the practical aspect of code time complexity prediction---the availability of annotated data.
By providing a fundamental framework, \texttt{TCProF}, for code time complexity prediction in low-resource settings,
we aim to lay the groundwork for future research in this domain.

\section{Related Works}
\subsection{SSL for Classification}
Data scarcity is a critical problem for various tasks.
More specifically, the problem occurs when there are only
a few labeled data even though there are tons of unlabeled data.
Generating labeled data is costly,
making research in these low-resource environments crucial.
Semi-supervised learning~(SSL) offers
an effective solution in such scenarios~\cite{SajjadiJT2016, XieDHLL2020, ChenZD2020, ChenHP2022, SohnBCZZ2020, ZhangWHW2021, WangCHH2022,ZouCZC23,Huang0Y0L23,NieDLWHZ24}.

One of the popularly used SSL techniques,
self-training is a learning mechanism that trains the student model with a few-shot labeled dataset~\cite{GengLLZJS19,BaoWCB20,ZhangB0CLXTCY21}
and then subsequently acts as a teacher model for generating pseudo-labels~\cite{Lee2013}.
Pseudo-labels are judged based on the predictions of the model for a given unlabeled data.
Co-training~\cite{BlumM98} is also the successful SSL mechanism
that simultaneously employs two networks.
Jointmatch~\cite{ZouC23} utilizes cross-labeling, inspired by co-training, that uses an additional loss based on pseudo-labels for more reliability
instead of augmenting pseudo-labels to the initial labeled dataset for additional training.
While this approach has been effective to some degree,
it lacks reliability as we cannot guarantee that the model generates `correct' pseudo-labels.
Recent approaches have incorporated symbolic modules
to enhance the reliability of pseudo-label generations or data augmentation~\cite{HahnCHLKH21,KimWOCH22}.

Likewise, data augmentation is also one of the fundamental methods that
is effective in the low-resource setting.
Data augmentation generates artificial data from the original dataset
without changing their labels.
Conventional data augmentations are synonym replacements, word insertion or deletion~\cite{WeiZ19}
More advanced methods involve Back-Translation~\cite{EdunovOAG18} and these days,
LLMs have gained popularity in generating various data but with accurate labels.
We utilize these insights to develop \texttt{TCProF},
an SSL framework for predicting code time complexity in low-resource environments.

\subsection{Code Time Complexity Prediction}
Computation of code time complexity has long been a theoretically undecidable problem
whereas classifying the code time complexity is a recently emerged problem.
\citet{SikkaSKUSZ20} first proposed this task and presented a labeled dataset with
five time complexity classes.
They propose a dataset named CorCoD,
composed of Java codes with $O(1)$, $O(\log N)$, $O(N)$, $O(N\log N)$ and $O(N^2)$ time complexity classes
and experimental results on the time complexity classification with baseline neural models.

Similar to CorCoD,
CODAIT\footnote{https://community.ibm.com/community/user/ai-datascience/blogs/sepideh-seifzadeh1/2021/10/05/ai-for-code-predict-code-complexity-using-ibms-cod}
tried to create good code embeddings by capturing manual features such 
as number of loops and breaks, and utilized graph-based representations
for predicting time complexities.
Afterward, \citet{MoudgalyaRCL23} proposed TasTy, consisting of Java and Python data, and \citet{BaikJHKHK24}
proposed CodeComplex, which consists of also Java and Python data with additional labels.
CodeComplex consists of seven different time complexity classes,
$O(N^3)$ and $O(2^N)$ in addition to those of CorCoD.

While time complexity prediction has been explored in these datasets,
the scarcity of labeled data remains a significant challenge.
Unlike runtime-based datasets from online judge platforms,
which can be influenced by hardware, input distributions,
and implementation-specific optimizations~\cite{IshimweNN21,ZhangWL23},
these datasets provide explicit theoretical complexity labels for the code snippets.
However, annotating such datasets requires expert knowledge, making them inherently low-resource.
We propose \texttt{TCProF}, the SSL framework designed to
alleviate this data scarcity challenge and enhance the accuracy of time complexity prediction.

\section{Methodology}\label{sec:method}

\subsection{Overview}\label{ssec:overview}
\texttt{TCProF} is a robust SSL framework designed for
code time complexity prediction in a low-resource setting illustrated in Fig.~\ref{fig:framework}.
\texttt{TCProF} comprises three primary components:
\paragraph{Augmentation module~($\mathbb{AUG}$):} This module employs augmentation involving
loop representation conversion~($LC$) and back-translation conversion~($BT$),
and a combined ensemble of these methods to enhance the diversity of augmented data.\\
\textbf{Training mechanism:} Utilizing a co-training approach,
\texttt{TCProF} trains two models simultaneously; one with original data and the other with augmented data.
This mitigates the error accumulation problem of self-training.\\
\textbf{Pseudo-label module:} Integrating our symbolic module~($\mathbb{Sym}$) with the classifier,
this module generates more precise pseudo-labels.

Algorithm~\ref{alg:overview} describes a detailed procedure of our framework.

\subsection{Self Training}\label{ssec:self-train}
In our experiments,
we have a dataset~$T$ for training,
a dataset~$V$ for validation and
a dataset~$E$ for evaluation.
we split $T$ into a labeled dataset~$L = \{l_1, l_2, \ldots, l_M\}$
of size~$M$ and an unlabeled dataset~$U = \{u_1, u_2, \ldots, u_N\}$ of size~$N$ for self-training:
\[
L = \{l\mid l = (d,c)\},~U = \{u\mid u = (d,\lambda) \}
\]
where $d$ is a code and $c$ is a complexity class for $d$.
$\lambda$ in $u$ represents that $u$ does not contain a complexity class for $d$.

For the first iteration, we train the baseline model~$B$ with $L$.
With the trained~$B$, we predict the complexity class~$c$ for $d$ in each~$u\in U$.
If the confidence score~$s_u(c)$ of $u$ for $c$ passes the pre-defined threshold~$\theta$,
we pseudo-label each $u$ with its corresponding label~$c_u$ and then add it to $L$:
\[
L' = L + \{u \mid u = (d, c_u) \text{ where $s_u(c_u) \ge \theta$}\}.
\]
For the next iterations afterward, we train $B$ with the updated labeled dataset~$L'$
and then pseudo-label the unlabeled data to update $L'$ again.

\begin{algorithm}[h]
\caption{Procedure for the SSL
framework~\texttt{TCProF}($\mathcal{L}$, $\mathcal{U}$,
$\mathbf{B}$, $\mathbf{B}_{\text{aug}}$, $\mathbb{Sym}$, $\mathbb{AUG}$).
Inputs include labeled dataset $\mathcal{L}$,
unlabeled dataset $\mathcal{U}$, baseline model $\mathbf{B}$,
co-training model $\mathbf{B}_{\text{aug}}$, symbolic module $\mathbb{Sym}$,
and augmentation module~$\mathbb{AUG}$.
The set of complexity classes is denoted as $\mathcal{C}$.}
\label{alg:overview}
\small
\begin{algorithmic}
\Function{Pseudo-label}{$\mathbf{B}$, $\mathbb{Sym}$, $\mathcal{U}$}
\Comment{Sections~\ref{ssec:model-pl},~\ref{ssec:symbolic-pl}}
    \State $\mathcal{U}' \gets \emptyset$ 
    \For{each $u = (d, \lambda) \in \mathcal{U}$}
        \State $c_u \gets \argmax_{c \in \mathcal{C}} \mathbf{B}(u, c)$
        \State $\mathcal{U}' \gets \mathcal{U}' \cup \{(d, c_u) : \mathbf{B}(u, c_u) \geq \theta\}$
        \State $\mathcal{U}' \gets \mathcal{U}' \cup \{(d, \mathbb{Sym}(d, u)) : \mathbf{B}(u, c_u) < \theta\}$
    \EndFor
    \State \Return $\mathcal{U}'$
\EndFunction

\Procedure{TimeComp}{$\mathcal{L}$, $\mathcal{U}$, $\mathbf{B}$, $\mathbf{B}_{\text{aug}}$, $\mathbb{Sym}$, $\mathbb{AUG}$}
    \State $\mathcal{L}_{\text{aug}} \gets \mathbb{AUG}(\mathcal{L})$ \Comment{Section~\ref{ssec:data-aug}}
    \For{$e \gets 1$ to $\text{epoch\_number}$}
        \If{self-train} \Comment{Section~\ref{ssec:self-train}}
            \State $\mathcal{L} \gets \mathcal{L} \cup \mathcal{L}_{\text{aug}}$
            \State $\mathbf{B} \gets \text{fine-tune}(\mathbf{B}, \mathcal{L})$
            \State $\mathcal{U}_{\text{pl}} \gets \Call{Pseudo-label}{\mathbf{B}, \mathbb{Sym}, \mathcal{U}}$
            \State $\mathcal{L} \gets \mathcal{L} \cup \mathcal{U}_{\text{pl}}$
        \ElsIf{co-train} \Comment{Section~\ref{ssec:co-training}}
            \State $\mathbf{B} \gets \text{fine-tune}(\mathbf{B}, \mathcal{L})$
            \State $\mathbf{B}_{\text{aug}} \gets \text{fine-tune}(\mathbf{B}_{\text{aug}}, \mathcal{L}_{\text{aug}})$
            \State $\mathcal{U}_{\text{pl}} \gets \Call{Pseudo-label}{\mathbf{B}, \mathbb{Sym}, \mathcal{U}}$
            \State $\mathcal{U}_{\text{aug-pl}} \gets \Call{Pseudo-label}{\mathbf{B}_{\text{aug}}, \mathbb{Sym}, \mathcal{U}}$
            \State $\mathcal{L} \gets \mathcal{L} \cup \mathcal{U}_{\text{aug-pl}}$
            \State $\mathcal{L}_{\text{aug}} \gets \mathcal{L}_{\text{aug}} \cup \mathcal{U}_{\text{pl}}$
        \EndIf
    \EndFor
\EndProcedure
\end{algorithmic}
\end{algorithm}

\begin{figure*}[h!]
    \centering
    \includegraphics[width=\textwidth]{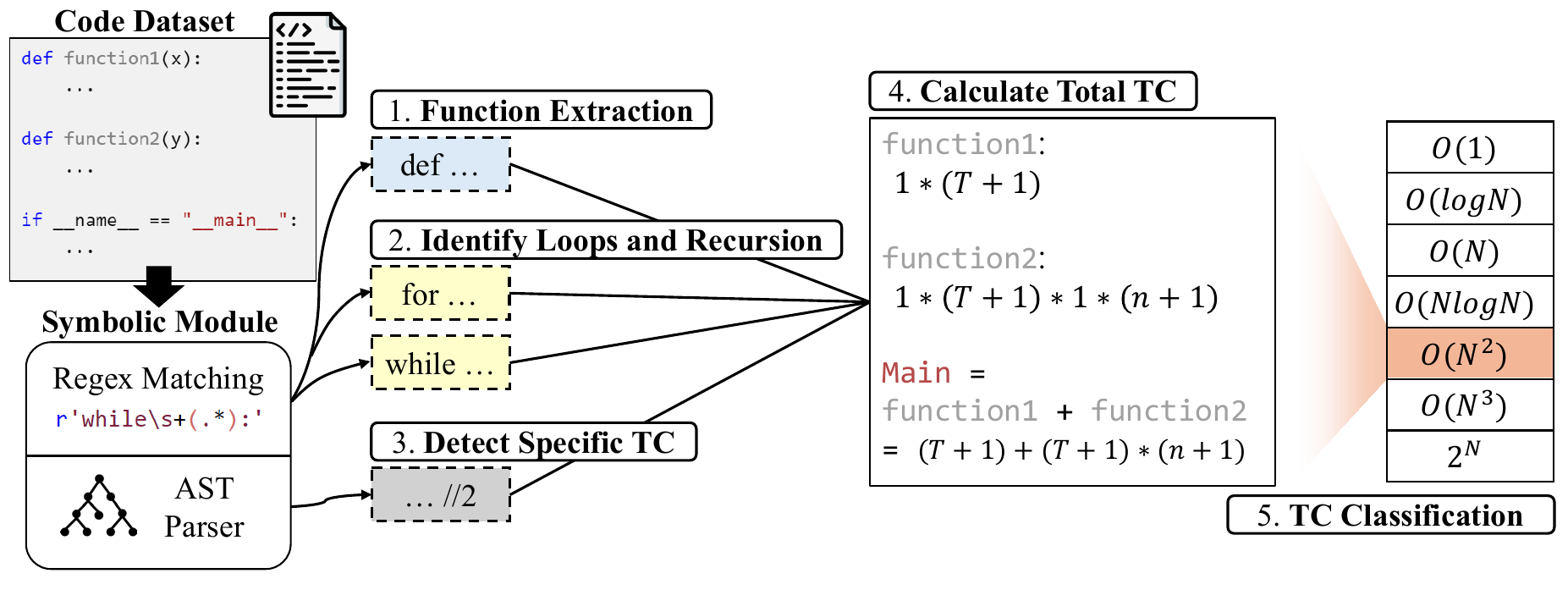}
    \caption{A procedural illustration of our symbolic module~$\mathbb{Sym}$.}
    \label{fig:symbolic-proc}
\end{figure*}

\subsection{Co-Training}\label{ssec:co-training}
We mitigate the error accumulation problem of self-training
by implementing co-training as our learning mechanism.
Our implementation of co-training involves two different models~$B$ and $B_{aug}$
which are trained with two different datasets~$L$ and $L_{aug}$.
Likewise to the self-training in Section~\ref{ssec:self-train},
$L$ is a labeled dataset of size $M$.
$L_{aug}$ is a labeled dataset of size $M$ generated by our data augmentation strategy in Section~\ref{ssec:data-aug}:
\[
\begin{split}
L_{aug} &= \{l_{aug} \mid l_{aug} = (d_{aug}, c), \text{ where }\\
&d_{aug} \text{ is augmented from } d \text{ in } l = (d, c)\}.
\end{split}
\]
Each model~$B$ and $B_{aug}$ is trained by $L$ and $L_{aug}$, respectively.
Following the pseudo-label strategy from Sections~\ref{ssec:model-pl} and~\ref{ssec:symbolic-pl},
two models then generate pseudo-labels and update labeled dataset $L'$ and $L'_{aug}$, respectively:
\[
\begin{split}
L' = L + &\{u\mid u = (d,c_u) \text{ where }\\
&u \text{ is pseudo-labeled by $B_{aug}$} \},
\end{split}
\]
\[
\begin{split}
L'_{aug} = L_{aug} + &\{u\mid u = (d,c_u) \text{ where }\\
&u \text{ is pseudo-labeled by $B$} \}.
\end{split}
\]

\subsection{Model-based Pseudo-Labels}\label{ssec:model-pl}
The conventional pseudo-labeling procedure is done by prediction of $B$.
The confidence score of unlabeled data with its corresponding class that $B$ outputs
is the one and only component for model-based pseudo-labels.
We can adjust the threshold value for more precise pseudo-labels.

\subsection{Symbolic Pseudo-Labels}\label{ssec:symbolic-pl}
We introduce a computational module,
designed to predict the temporal complexity
of source code through symbolic analysis, employing the computational results as
pseudo-labels to enhance the accuracy of predictions. 
We have named this module, the \textbf{symbolic module}~$\mathbb{Sym}$.
The symbolic module~$\mathbb{Sym}$ is designed to predict the time complexity of code by identifying specific
patterns and structural elements, without relying on neural networks.
This approach effectively 
complements language models that, by their nature, do not account for the hierarchical 
organization inherent in programming code~\citep{AllamanisBK18, ChenHLMMTM21, ZhangWZ0J22}.
The module primarily employs Regular Expressions~(Regex) and
Abstract Syntax Trees~(ASTs) as essential tools for a detailed analysis of the source code.
This aims to describe the time complexity of iterations, function calls, recursive and iterative calls,
and additional relevant constructs, thereby facilitating a comprehensive understanding 
of the code's structure and execution flow.
Our module outputs a formula consisting of the above components and regarding the size of
inputs in the source code, subsequently aggregating these elements to categorize 
the time complexity class of the source code.

Our approach is systematically organized 
into five distinct phases and is illustrated in Figure~\ref{fig:symbolic-proc}:
    
\paragraph{Function Extraction:} We employ Regex to identify and extract function definitions from source code.
Each function forms the basic unit of analysis as an independent code block.
If the source code does not contain functions, this step is omitted.\\
\textbf{Identify Loops and Recursions:} We utilize Regex to identify the presence of 
repetitive statements (such as for loops and while loops) and recursive functions to determine 
the frequency of loop iterations and function calls.\\
\textbf{Detect Specific Time Complexity~(TC):} We utilize ASTs
to detect operations that modify the input size.
By identifying the presence of sorting or binary 
traversal, we classify source codes according to their
time complexities such as $O(\log N)$ and $O(N\log N)$.\\
\textbf{Calculate Total TC:} The final time complexity is calculated
by summing the complexity of each iteration and function call,
based on the patterns identified using Regex, and the code structure analyzed by ASTs.
In this step, we calculate the overall complexity,
ensuring a robust analysis of recursive relationships within the source code.\\
\textbf{TC Classification:} The calculated time complexity is categorized into 
the pre-defined complexity classes.
These assigned classes are utilized as pseudo-labels.

In Appendix~\ref{app:symbolic}, we present Figure~\ref{fig:sym_fig} as a running example of the symbolic pseudo-labeling.

\subsection{Merge Pseudo-Labeling}
We have two pseudo-label modules based on model confidence and the symbolic module.
Illustrated in Figure~\ref{fig:framework} and Algorithm~\ref{alg:overview},
we first pseudo-label the unlabeled data by the model confidence.
Then, we use the symbolic module to pseudo-label the unlabeled data that failed pseudo-labeling by the model confidence.

\subsection{Data Augmentation}\label{ssec:data-aug}
We introduce a data augmentation module~$\mathbb{AUG}$, designed to complement the lack of labeled data.
Our augmentation strategies leverage the ChatGPT API,
specifically employing the \texttt{gpt-3.5-turbo-0125} model,
to augment our experiments' CorCoD and CodeComplex datasets.
The objective is to create precise augmentations that respect the intrinsic properties of the code,
ensuring semantic integrity while introducing syntactic variability.
We ensure that the augmented code snippets are free from syntactic errors, reinforcing their reliability for further analysis.
We present two augmentation methods specifically designed to augment code snippets that preserve the original time complexity:

\paragraph{Back-Translation~(BT):}
This method involves translating a code snippet into another programming language and then
back to the original language to maintain its semantic essence.
For instance, Java code snippets are translated into Python and then back into Java.
This process gives syntactic variation while retaining the context.\\
\textbf{Loop-Conversion~(LC)}:
Loop structures are the primary components that determine the time complexity of code snippets.
This technique modifies the loop structures to different but semantically equivalent forms.
Using regular expressions, we filter codes containing ``for'' or ``while'' loops and then
convert these loops by employing For2While and While2For transformation rules, preserving the original logic of the code snippet.
For instance, a while loop can be converted into a for loop~(While2For) and vice versa~(For2While), depending on the context.
If the original code snippet contains both for and while loops,
we leverage While2for and For2while respectively.

These augmentation methods are detailed in the prompts listed in Appendix~\ref{app:aug-prompt}
and are integrated into \texttt{TCProF} to enhance the robustness.
The augmented data from BT and LC methods supplement the initial labeled data in three distinct experimental configurations:
1) $\mathbb{AUG}_{BT}$ incorporates back-translation augmented data,
2) $\mathbb{AUG}_{LC}$ incorporates loop-conversion augmented data, and
3) $\mathbb{AUG}_{BT+LC}$ combines both back-translation and loop-conversion augmented data.

There are two experimental setups for $\mathbb{AUG}_{LC}$ as there are code snippets without any loop structures:
1) $\mathbb{AUG}_{LC\_Natural}$ uses naturally sampled data as the initial labeled data without specific preconditions and
2) $\mathbb{AUG}_{LC\_Artificial}$ selects initial labeled data specifically containing loop structures to
maximize the use of augmented data.

\begin{table*}[ht]
\centering
\resizebox{\textwidth}{!}{%
\begin{tabular}{lcccccc}
\toprule
               &  \multicolumn{2}{c}{CodeComplex~(Java)}  &  \multicolumn{2}{c}{CodeComplex~(Python)}  &        \multicolumn{2}{c}{CorCoD}           \\
               &          5          &         10         &          5           &         10          &           5           &         10          \\
\hline
\multicolumn{7}{l}{SSL Baselines} \\
\hline
ST~(UniXcoder)             &15.44\tiny$\pm$ 6.30          &31.77\tiny$\pm$ 0.81          &26.02\tiny$\pm$11.72         &40.98\tiny$\pm$ 1.14         &37.89\tiny$\pm$ 6.57         &45.61\tiny$\pm$11.59 \\
ST~(CodeT5+)               &18.87\tiny$\pm$ 3.21          &31.69\tiny$\pm$14.00          &28.76\tiny$\pm$17.47         &38.32\tiny$\pm$18.82         &35.79\tiny$\pm$ 6.90         &45.26\tiny$\pm$ 6.32 \\
JointMatch~(CodeT5+)       &14.62\tiny$\pm$ 4.26          &24.68\tiny$\pm$ 2.60          &20.97\tiny$\pm$ 3.90         &21.04\tiny$\pm$ 4.33         &36.49\tiny$\pm$ 4.98         &42.11\tiny$\pm$ 6.57 \\
JointMatch~(UniXcoder)     & 9.62\tiny$\pm$ 5.23          &19.39\tiny$\pm$ 3.86          &14.68\tiny$\pm$11.78         &20.76\tiny$\pm$10.46         &35.44\tiny$\pm$ 6.85         &48.42\tiny$\pm$ 8.22 \\
\midrule                                        
\midrule                                        
\texttt{TCProF}(CodeT5+)   &38.63\tiny$\pm$ 1.32          &41.98\tiny$\pm$ 2.92          &  44.74\tiny$\pm$ 3.26        &59.29\tiny$\pm$ 3.71          &50.53\tiny$\pm$ 0.86          &51.93\tiny$\pm$ 3.88 \\
\texttt{TCProF}(UniXcoder) &\textbf{52.50}\tiny$\pm$ 1.56 &\textbf{53.85}\tiny$\pm$ 3.63 &\textbf{54.64}\tiny$\pm$ 3.77 &\textbf{70.29}\tiny$\pm$ 2.06 &\textbf{55.44}\tiny$\pm$ 0.99 &\textbf{63.16}\tiny$\pm$ 2.27 \\
\bottomrule
\end{tabular}
}
\caption{Accuracy performance of SSL baselines and \texttt{TCProF}.
ST refers to Self-Training.
\texttt{TCProF}(CodeT5+) represents \texttt{TCProF} implemented to the baseline~CodeT5+
and vice versa to UniXcoder.
The scores are averaged from three runs with different seeds.
We report the full result in Table~\ref{tbl:baseline-comparison-acc-full}.}
\label{tbl:baseline-comparison-acc-part}
\end{table*}

These configurations for $\mathbb{AUG}_{LC}$ are also applied
to the combined augmentation strategy~$\mathbb{AUG}_{BT+LC}$,
with corresponding $\mathbb{AUG}_{BT+LC\_Natural}$
and $\mathbb{AUG}_{BT+LC\_Artificial}$ settings.
Table~\ref{tbl:ablation-studies} employs $\mathbb{AUG}_{BT+LC\_Natural}$
as the baseline for the $\mathbb{AUG}$ setup as
it is more natural and common compared to $\mathbb{AUG}_{BT+LC\_Artificial}$.
An extensive analysis of augmentation strategies is provided in Appendix~\ref{app:aug-analysis}.

\subsection{Implementation Details}
We use UniXcoder~\cite{GuoLDW0022} and CodeT5+~\cite{WangLGB0H23}
as our baselines for \texttt{TCProF}.
We assign the batch of size~7,
the number of epochs as 20,
and set the learning rate of co-training as \texttt{1e-5} and \texttt{2e-6}.
We use \texttt{1e-5} learning rate for self-training.
Our confidence score threshold~$\theta$ is 0.7.
We conduct experiments using the NVIDIA RTX 3090 for training \texttt{TCProF}.

\section{Experimental Setup}
We use CodeComplex~\cite{BaikJHKHK24} and CorCoD~\cite{SikkaSKUSZ20} datasets.
CodeComplex consists of 4,900 Java and 4,900 Python codes, and
CorCoD consists of 929 Java codes.
We follow \citet{BaikJHKHK24} to split train and test data.
For low-resource settings,
we perform 5- and 10-shot experiments where we pick 5 and 10 data for each label
from the train dataset and use the remaining train dataset as an unlabeled dataset for pseudo-labels, respectively.

Code time complexity prediction is a recent code-related task and we are the first to perform
the task in low-resource scenarios.
We use four pre-trained code language models,
CodeBERT~\cite{FengGTDFGS0LJZ20}, GraphCodeBERT~\cite{GuoRLFT0ZDSFTDC21},
UniXcoder~\cite{GuoLDW0022}, and CodeT5+~\cite{WangLGB0H23}
as our baseline models.

We also include JointMatch as an SSL baseline for comparison.
Recently, \citet{ZouC23} proposed JointMatch for a state-of-the-art SSL framework for low-resource settings,
where the authors employ cross-labeling to enhance pseudo-labeling.
While JointMatch is effective for text classification tasks in low-resource settings,
we analyze whether it is effective for code time complexity prediction in low-resource settings
and compare the performance with our framework~\texttt{TCProF}.

\section{Results and Analysis}
We present our main results in Table~\ref{tbl:baseline-comparison-acc-part},
indicating that our model outperforms baselines.
In Section~\ref{ssec:baseline-comparison}, we suggest possible intuitions that
can be derived from the performance result of Table~\ref{tbl:baseline-comparison-acc-part}.
We demonstrate ablation studies in Section~\ref{ssec:ablation}
and provide error analyses in Section~\ref{ssec:error-analysis}.
Then, for the extensive analyses of \texttt{TCProF},
we show how TCProF approaches the fully fine-tuned models in Section~\ref{ssec:full-comparison}
and provide comparisons with commercial LLMs in Section~\ref{ssec:llm-comparison}.

\subsection{Comparison with Baselines}\label{ssec:baseline-comparison}
Table~\ref{tbl:baseline-comparison-acc-part} presents the
Self-Training~(ST) results for the base models, UniXcoder and CodeT5+.
The table illustrates that the average accuracy and standard deviation scores
for ST of base models, which serve as baselines, are generally low.
Specifically, on the CodeComplex datasets, the accuracy scores of these baselines are below 40\%,
except for ST~(UniXcoder) on CodeComplex~(Python).
Furthermore, the high standard deviation scores
indicate that the performance of these models is unstable and unreliable.

\begin{table*}[ht]
\centering
\resizebox{\textwidth}{!}{%
\begin{tabular}{lcccccc}
\hline
                                                                      &  \multicolumn{2}{c}{CodeComplex~(Java)}  &  \multicolumn{2}{c}{CodeComplex~(Python)}  &        \multicolumn{2}{c}{CorCoD}           \\
                                                                      &          5          &         10         &          5           &         10          &           5           &         10          \\
\hline
\multicolumn{7}{c}{CodeT5+} \\
\hline
\hspace{4pt}+ \hspace{2pt}$\mathbb{AUG}$                      &26.85\tiny$\pm$ 2.4   &35.86\tiny$\pm$ 4.53 &  33.94\tiny$\pm$ 3.82 &57.44\tiny$\pm$ 1.17  &36.14\tiny$\pm$ 4.98    &43.85\tiny$\pm$ 6.34 \\
\hspace{4pt}+ \hspace{2pt}$\mathbb{Sym}$                              &21.92\tiny$\pm$ 2.54  &35.50\tiny$\pm$11.18 &  32.10\tiny$\pm$ 6.78 &46.38\tiny$\pm$ 6.58  &36.84\tiny$\pm$ 1.72    &42.81\tiny$\pm$ 2.76 \\
\hspace{4pt}+ \hspace{2pt}$\mathbb{Sym}$ + $\mathbb{AUG}$     &37.88\tiny$\pm$ 1.65  &38.26\tiny$\pm$ 2.40 &  41.80\tiny$\pm$ 1.49 &58.88\tiny$\pm$12.37  &37.89\tiny$\pm$ 2.58    &44.21\tiny$\pm$ 2.27 \\
\midrule
\hspace{4pt}\texttt{TCProF}(CodeT5+)                                  &38.63\tiny$\pm$ 1.32  &41.98\tiny$\pm$ 2.92 &  44.74\tiny$\pm$ 3.26 &59.29\tiny$\pm$ 3.71  &50.53\tiny$\pm$ 0.86    &51.93\tiny$\pm$ 3.88 \\
\hline
\hline
\multicolumn{7}{c}{UnixCoder} \\
\hline
\hspace{4pt}+ $\mathbb{AUG}$                      &34.22\tiny$\pm$12.32  &39.00\tiny$\pm$ 6.64 &51.63\tiny$\pm$12.42   &63.04\tiny$\pm$ 8.75  &44.21\tiny$\pm$ 3.16    &51.57\tiny$\pm$ 5.57 \\
\hspace{4pt}+ $\mathbb{Sym}$                              &41.76\tiny$\pm$ 2.62  &45.34\tiny$\pm$ 2.91 &39.82\tiny$\pm$ 5.13   &51.02\tiny$\pm$ 2.14  &50.88\tiny$\pm$ 8.48    &54.04\tiny$\pm$ 6.56 \\
\hspace{4pt}+ $\mathbb{Sym}$ + $\mathbb{AUG}$     &43.70\tiny$\pm$ 2.00  &45.49\tiny$\pm$ 1.10 &54.03\tiny$\pm$ 5.52   &67.55\tiny$\pm$ 1.06  &50.18\tiny$\pm$ 4.24    &58.60\tiny$\pm$ 4.73 \\
\midrule
\hspace{4pt}\texttt{TCProF}(UniXcoder)                    &\textbf{52.50}\tiny$\pm$ 1.56  &\textbf{53.85}\tiny$\pm$ 3.63 &\textbf{54.64}\tiny$\pm$ 3.77   &\textbf{70.29}\tiny$\pm$ 2.06  &\textbf{55.44}\tiny$\pm$ 0.99    &\textbf{63.16}\tiny$\pm$ 2.27 \\
\bottomrule
\end{tabular}
}
\caption{Ablation studies of \texttt{TCProF}(CodeT5+) and \texttt{TCProF}(UniXcoder) accuracy~(\%) performance.
$\mathbb{AUG}$ represents $\mathbb{AUG}_{BT+LC}$.
The full result is in Table~\ref{tbl:baseline-comparison-acc-full} and extensive augmentation result is in Table~\ref{tbl:extensive-aug-acc}.}
\label{tbl:ablation-studies}
\end{table*}

Our augmentation, symbolic pseudo-labeling, and co-training strategies mitigate this problem.
We pick UniXcoder and CodeT5+ as the baseline model to implement our strategies\footnote{Appendix~\ref{app:baseline-selection}
demonstrates the detailed analysis of the comparison between UniXcoder and CodeT5+.}.
\texttt{TCProF} outperforms the baselines
illustrated in Table~\ref{tbl:baseline-comparison-acc-part}.
Especially, \texttt{TCProF}(UniXcoder) accomplishes 64.81\% improvements on average compared to the best performance of baselines.
We discuss the analysis of our strategies in Section~\ref{ssec:ablation} as ablation studies.

We also include experiments on JointMatch, a state-of-the-art SSL baseline, to compare with \texttt{TCProF}.
While JointMatch is the state-of-the-art approach for text classification in low-resource settings,
the case is quite different in a code time complexity prediction.
Table~\ref{tbl:baseline-comparison-acc-part} indicates that JointMatch is generally less effective
than self-training baselines and \texttt{TCProF}(UniXcoder) outperforms JointMatch by 131.09\% on average.
This is because cross-labeling of JointMatch depends on unlabeled data to alleviate
the pseudo-label noise and the strategy is not quite applicable to code time complexity prediction data.
We provide a detailed analysis in Appendix~\ref{app:compare-jointmatch}
illustrating that \texttt{TCProF} is more appropriate than the cross-labeling of JointMatch.

\subsection{Ablation Studies}\label{ssec:ablation}
In our ablation studies, delineated within Table~\ref{tbl:ablation-studies},
we evaluate the impact of each module within \texttt{TCProF},
as introduced in Figure~\ref{fig:framework} and Section~\ref{sec:method}.
Our framework integrates the augmentation module~$\mathbb{AUG}$,
the symbolic module~$\mathbb{Sym}$, and
implements both self-training and co-training strategies
for code time complexity prediction in low-resource settings.
The underlying hypothesis points that each component incrementally improves upon
basic self-training baseline models.
The empirical result from Table~\ref{tbl:ablation-studies} confirms our hypothesis,
demonstrating performance enhancements in our baselines, CodeT5+ and UniXcoder,
as additional modules are integrated.

Particularly, the interaction between
$\mathbb{AUG}$ and $\mathbb{Sym}$ shows a clear synergistic effect.
While $\mathbb{AUG}$ alone significantly boosts performance compared to the self-training baselines,
it mostly shows high standard deviation scores, which are then relaxed when applied with $\mathbb{Sym}$.
This is especially noticeable in CodeComplex~(Python).
$\mathbb{AUG}$ is effective considering its enhancements from the performance of self-training baselines in Table~\ref{tbl:baseline-comparison-acc-part},
but it also shows high standard deviation scores, mostly over 10\%.
However, this problem is substantially mitigated when
$\mathbb{AUG}$ is combined with $\mathbb{Sym}$,
reducing the average standard deviation to 5\%.

This enhancement is further developed when we implement the co-training strategy,
solidifying not only the accuracy but also reducing the standard deviation.
These observations are particularly pronounced in Python datasets rather than Java datasets.
The variance between Python and Java can primarily be attributed to the differential effectiveness of
$\mathbb{AUG}$ across these programming languages.
The less stringent syntax of Python compared to Java allows greater variability in augmented Python code snippets,
contributing to this performance boost.
This phenomenon is further supported by Tables~\ref{tbl:extensive-aug-acc} and~\ref{tbl:extensive-aug-f1}
in Appendix~\ref{app:aug-analysis}.
Despite these differences, a closer analysis of the F1-scores in Table~\ref{tbl:baseline-comparison-f1}
shows that both languages exhibit substantial improvements, each around 50\%.

Furthermore, implemented all together with a co-training strategy,
\texttt{TCProF} strengthens the performance
both in accuracy and standard deviation.
We also provide detailed analysis on F1-scores in Appendix~\ref{app:full-result}.

\begin{figure}[ht]
    \centering
    \includegraphics[width=\columnwidth]{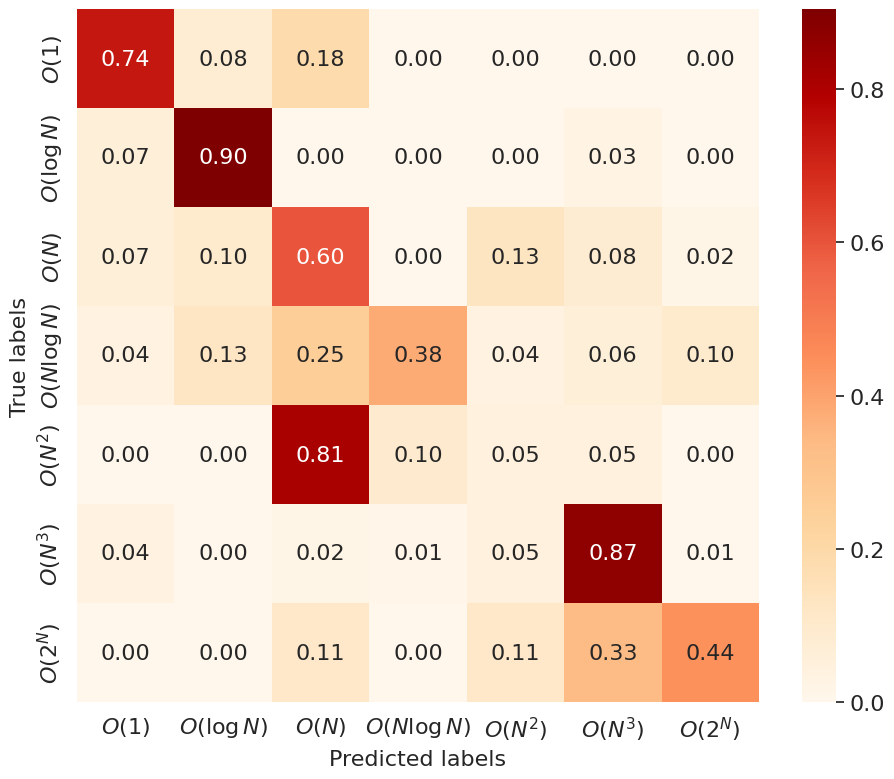}
    \caption{Confusion matrix of ours for CodeComplex~(Python) 10-shot.}
    \label{fig:cross-unixcoder-confusion}
\end{figure}

\subsection{Error Analysis}\label{ssec:error-analysis}
Furthermore,
we examine the types of errors to which our method is particularly vulnerable
for extensive analysis.
Figure~\ref{fig:cross-unixcoder-confusion} illustrates that our approach is
especially weak on $O(N^2)$ class.
We notice that the codes with this class are mostly predicted as $O(N)$.
Codes in both classes involve loops and usually the only difference is the depth of the loops.
However, precisely, the core factor that makes the difference is the loops involving the input length.
For instance, codes in $O(N)$~class can contain multiple-depth loop statements where only a single loop statement involves
the length of the input.
The model needs to precisely determine whether the iteration number of each loop is proportional to the input size
for the correct prediction.
Likewise, the class~$O(N log N)$ involving both loops linear and logarithmic to the input size and
the class~$O(2^N)$ involving exponential iterations are also relatively erroneous compared to the other classes.
We present examples of code instances, demonstrating the errors in Appendix~\ref{app:error-analysis}.

\subsection{Comparison to the full-train result}\label{ssec:full-comparison}
\texttt{TCProF} demonstrates a promising result in a few-shot settings,
improving the baselines on a large scale.
We conduct further experiments on comparing the best 10-shot performance of \texttt{TCProF}
to the baselines trained with the whole train datasets in Table~\ref{tbl:baseline-full-acc}.
For CodeComplex datasets, \texttt{TCProF}(UniXcoder) exceeds the accuracy of all the baselines except for CodeT5+
and the performance gap with CodeT5+ is small.
Take CodeComplex~(Python) for instance,
\texttt{TCProF}(UniXcoder) achieves 70.29\% accuracy and
the accuracy of CodeT5+ fine-tuned with the full train dataset
is 72.88\%.
Our approach catches up to 96.45\%.

\begin{table}[ht]
\centering
\resizebox{\columnwidth}{!}{
\begin{tabular}{lccc}
\hline
\multicolumn{1}{c}{\multirow{2}{*}{FULL}}      &            \multicolumn{2}{c}{CodeComplex}                    & \multirow{2}{*}{CorCoD}        \\
                                               &        Java                  &        Python                  &                                \\\hline\hline
                                               &         Acc.                 &         Acc.                   &         Acc.                   \\
\midrule                               
CodeBERT                                       &39.82\tiny$\pm$ 6.90          &66.39\tiny$\pm$  1.43           &74.74\tiny$\pm$  1.72           \\
GraphCodeBERT                                  &46.46\tiny$\pm$ 2.18          &67.83\tiny$\pm$  8.36           &72.98\tiny$\pm$  1.79           \\
UniXcoder                                      &46.76\tiny$\pm$ 1.91          &68.44\tiny$\pm$  4.66           &\textbf{77.54}\tiny$\pm$  1.31  \\
CodeT5+                                        &\textbf{55.11}\tiny$\pm$ 2.18 &\textbf{72.88}\tiny$\pm$  1.95  &75.79\tiny$\pm$  2.27           \\
\bottomrule
\end{tabular}
}
\caption{Accuracy of base models trained with the full-train dataset.}
\label{tbl:baseline-full-acc}
\end{table}

Appendix~\ref{app:full-baseline-finetune} presents the full results of accuracy and F1 performance in Table~\ref{tbl:baseline-comparison-full}.
Overall to all the datasets,
\texttt{TCProF} catches up to 91\% and 94\% for accuracy and F1-scores, respectively,
compared to the base models fine-tuned with the full train dataset.
This indicates that \texttt{TCProF}, a framework designed for low-resource settings,
is highly effective in achieving competitive performance even with limited labeled data.
By leveraging co-training, symbolic pseudo-labels, and data augmentation, \texttt{TCProF} enhances model generalization and robustness,
making it a viable alternative to fully supervised approaches, particularly in scenarios where the amount of labeled data is scarce or costly to obtain.
We also present Figures~\ref{fig:full-acc-graph} and~\ref{fig:full-f1-graph} in the appendix for clear visualization of the results.

\begin{table}[ht]
\centering
\resizebox{\columnwidth}{!}{
\begin{tabular}{lccc}
\toprule
                                &    \multicolumn{2}{c}{CodeComplex}   & \multirow{2}{*}{CorCoD}       \\
                                &        Java      &        Python     &                               \\\hline\hline
                                &         Acc.     &         Acc.      &         Acc.                  \\\midrule
Gemini-pro                      &49.54             &   31.05           &    61.91                      \\
GPT-3.5                          &62.15             &   32.55           &    69.42                      \\
GPT-4                            &\textbf{64.01}    &   53.04           &   \textbf{78.86}              \\
\midrule              
\texttt{TCProF}(UniXcoder)      &53.85             &\textbf{70.29}     &  63.16                        \\
\bottomrule
\end{tabular}
}
\caption{Comparison with the performance of LLM for Java and Python datasets of CodeComplex, and CorCod dataset.}
\label{tbl:llm-comparison-acc}
\end{table}

\subsection{Comparison with LLMs}\label{ssec:llm-comparison}
LLMs are known to be effective in general NLP tasks and
we extend the analysis into comparing our performance with LLMs.
We evaluate the LLMs in a 5-shot in-context learning setting
by the prompt demonstrated in Appendix~\ref{app:llm-prompt}.
Our baseline LLMs are Gemini-pro~\cite{Gemini23}, GPT3.5, and GPT-4~\cite{gpt-report}.
The results of Table~\ref{tbl:llm-comparison-acc}
indicates that \texttt{TCProF} shows competitive performance to these commercial LLMs.
These LLMs perform much better than CodeT5+ and UniXcoder fine-tuned with the full train dataset
for CodeComplex~(Java) and CorCoD referring to Table~\ref{tbl:baseline-full-acc}.
As \texttt{TCProF} takes CodeT5+ and UniXcoder as the baseline models,
competitive performance of \texttt{TCProF} compared to LLMs is remarkable.

Furthermore, \texttt{TCProF} performs better than any of these LLMs for CodeComplex~(Python).
In Appendix~\ref{app:llm-comparison}, we also present Table~\ref{tbl:llm-comparison}
that includes both the accuracy and F1 performance of LLMs.

\section{Conclusion}\label{sec:conclusion}
Code time complexity prediction remains a largely unexplored yet critical task,
and our framework, \texttt{TCProF}, is the first attempt
to tackle this challenge within low-resource environments.
We have developed \texttt{TCProF} as a robust SSL framework
for predicting code time complexity, showing through comprehensive analyses
that it significantly improves upon the performance of baseline models.
As we reflect on the capabilities of \texttt{TCProF},
it is important to recognize that our current focus has been on establishing
an effective framework that performs well in constrained environments.
Looking ahead, expanding the generalization capabilities of \texttt{TCProF}
to accommodate a wider range of programming languages is a vital next step.
We believe \texttt{TCProF} will serve as a significant milestone,
propelling forward the research in code time complexity prediction, especially in low-resource settings.

\section*{Limitation}\label{sec:limitation}
\paragraph{Application Scope}
\texttt{TCProF} is specifically designed for few-shot settings and thus,
its application would not be effective in fine-tuning with the full dataset.
One possible exploration is employing the augmentation module of \texttt{TCProF}
for fine-tuning the full dataset.
We provide the analyses in Appendix~\ref{app:aug-analysis}.
Another limitation of \texttt{TCProF} is the space complexity prediction.
This is also due to the absence of datasets involving the space complexity.
Thus, we will be happy to apply and develop our framework for the space complexity prediction
when the datasets are released.

\paragraph{Framework Adaptability to Zero-Shots}
As our framework targets few-shot settings where `several' data instances for each class are provided,
We do not provide initial configuration for zero-shot settings.
However, we can adapt the zero-shot setting by altering the structure of \texttt{TCProF}
to incorporate the symbolic module for generating initial training data with pseudo-labels.
It is essential to clarify that our objective is not zero-shot learning but few-shot learning.

\paragraph{Dynamic Calculation}
\texttt{TCProF} is focused on classifying time complexity into seven discrete classes,
rather than dynamically calculating time complexities across a continuous range.
We are eager to develop further modules and implement generative models
for targeting the time complexity `calculation',
extending from the time complexity `prediction'.

\section*{Acknowledgment}
This research was supported by the NRF grant (RS-2023-00208094) and the AI Graduate School Program (RS-2020-II201361) funded by the Korean government~(MSIT).

\bibliography{arxiv}

\begin{thebibliography}{35}
\providecommand{\natexlab}[1]{#1}

\bibitem[{Allamanis et~al.(2018)Allamanis, Brockschmidt, and Khademi}]{AllamanisBK18}
Miltiadis Allamanis, Marc Brockschmidt, and Mahmoud Khademi. 2018.
\newblock Learning to represent programs with graphs.
\newblock In \emph{6th International Conference on Learning Representations, {ICLR}}.

\bibitem[{Anil and et~al.(2023)}]{Gemini23}
Rohan Anil and et~al. 2023.
\newblock Gemini: {A} family of highly capable multimodal models.
\newblock \emph{preprint}, arXiv: 2312.11805.

\bibitem[{Asperti(2008)}]{Asperti08}
Andrea Asperti. 2008.
\newblock The intensional content of {R}ice's theorem.
\newblock In \emph{Proceedings of the 35th {ACM} {SIGPLAN-SIGACT} Symposium on Principles of Programming Languages, {POPL}}, pages 113--119.

\bibitem[{Baik et~al.(2024)Baik, Hahn, Kim, Aditi, Jeon, Han, and Ko}]{BaikJHKHK24}
Seung{-}Yeop Baik, Joonghyuk Hahn, Jungin Kim, Aditi, Mingi Jeon, Yo{-}Sub Han, and Sang{-}Ki Ko. 2024.
\newblock {CodeComplex}: Dataset for worst-case time complexity prediction.
\newblock \emph{preprint}, arXiv: 2401.08719.

\bibitem[{Bao et~al.(2020)Bao, Wu, Chang, and Barzilay}]{BaoWCB20}
Yujia Bao, Menghua Wu, Shiyu Chang, and Regina Barzilay. 2020.
\newblock Few-shot text classification with distributional signatures.
\newblock In \emph{8th International Conference on Learning Representations, {ICLR}}.

\bibitem[{Blum and Mitchell(1998)}]{BlumM98}
Avrim Blum and Tom~M. Mitchell. 1998.
\newblock Combining labeled and unlabeled data with co-training.
\newblock In \emph{Proceedings of the Eleventh Annual Conference on Computational Learning Theory, {COLT}}, pages 92--100.

\bibitem[{Chen et~al.(2022)Chen, Han, and Poria}]{ChenHP2022}
Hui Chen, Wei Han, and Soujanya Poria. 2022.
\newblock {SAT:} improving semi-supervised text classification with simple instance-adaptive self-training.
\newblock In \emph{Findings of the Association for Computational Linguistics: {EMNLP}}, pages 6141--6146.

\bibitem[{Chen et~al.(2020)Chen, Yang, and Yang}]{ChenZD2020}
Jiaao Chen, Zichao Yang, and Diyi Yang. 2020.
\newblock Mixtext: Linguistically-informed interpolation of hidden space for semi-supervised text classification.
\newblock In \emph{Proceedings of the 58th Annual Meeting of the Association for Computational Linguistics, {ACL}}, pages 2147--2157.

\bibitem[{Chen et~al.(2021)Chen, Hellendoorn, Lamblin, Maniatis, Manzagol, Tarlow, and Moitra}]{ChenHLMMTM21}
Zimin Chen, Vincent~J. Hellendoorn, Pascal Lamblin, Petros Maniatis, Pierre{-}Antoine Manzagol, Daniel Tarlow, and Subhodeep Moitra. 2021.
\newblock {PLUR:} {A} unifying, graph-based view of program learning, understanding, and repair.
\newblock In \emph{Advances in Neural Information Processing Systems 34}, pages 23089--23101.

\bibitem[{Edunov et~al.(2018)Edunov, Ott, Auli, and Grangier}]{EdunovOAG18}
Sergey Edunov, Myle Ott, Michael Auli, and David Grangier. 2018.
\newblock Understanding back-translation at scale.
\newblock In \emph{Proceedings of the 2018 Conference on Empirical Methods in Natural Language Processing, {EMNLP}}, pages 489--500.

\bibitem[{Feng et~al.(2020)Feng, Guo, Tang, Duan, Feng, Gong, Shou, Qin, Liu, Jiang, and Zhou}]{FengGTDFGS0LJZ20}
Zhangyin Feng, Daya Guo, Duyu Tang, Nan Duan, Xiaocheng Feng, Ming Gong, Linjun Shou, Bing Qin, Ting Liu, Daxin Jiang, and Ming Zhou. 2020.
\newblock Code{BERT}: {A} pre-trained model for programming and natural languages.
\newblock In \emph{Findings of the Association for Computational Linguistics: {EMNLP}}, pages 1536--1547.

\bibitem[{Geng et~al.(2019)Geng, Li, Li, Zhu, Jian, and Sun}]{GengLLZJS19}
Ruiying Geng, Binhua Li, Yongbin Li, Xiaodan Zhu, Ping Jian, and Jian Sun. 2019.
\newblock Induction networks for few-shot text classification.
\newblock In \emph{Proceedings of the 2019 Conference on Empirical Methods in Natural Language Processing and the 9th International Joint Conference on Natural Language Processing, {EMNLP-IJCNLP}}, pages 3902--3911.

\bibitem[{Guo et~al.(2022)Guo, Lu, Duan, Wang, Zhou, and Yin}]{GuoLDW0022}
Daya Guo, Shuai Lu, Nan Duan, Yanlin Wang, Ming Zhou, and Jian Yin. 2022.
\newblock Uni{X}coder: Unified cross-modal pre-training for code representation.
\newblock In \emph{Proceedings of the 60th Annual Meeting of the Association for Computational Linguistics, {ACL}}, pages 7212--7225.

\bibitem[{Guo et~al.(2021)Guo, Ren, Lu, Feng, Tang, Liu, Zhou, Duan, Svyatkovskiy, Fu, Tufano, Deng, Clement, Drain, Sundaresan, Yin, Jiang, and Zhou}]{GuoRLFT0ZDSFTDC21}
Daya Guo, Shuo Ren, Shuai Lu, Zhangyin Feng, Duyu Tang, Shujie Liu, Long Zhou, Nan Duan, Alexey Svyatkovskiy, Shengyu Fu, Michele Tufano, Shao~Kun Deng, Colin~B. Clement, Dawn Drain, Neel Sundaresan, Jian Yin, Daxin Jiang, and Ming Zhou. 2021.
\newblock Graph{C}ode{BERT}: Pre-training code representations with data flow.
\newblock In \emph{9th International Conference on Learning Representations, {ICLR}}.

\bibitem[{Hahn et~al.(2021)Hahn, Cheon, Han, Lee, Kim, and Han}]{HahnCHLKH21}
Joonghyuk Hahn, Hyunjoon Cheon, Kyuyeol Han, Cheongjae Lee, Junseok Kim, and Yo{-}Sub Han. 2021.
\newblock Self-training using rules of grammar for few-shot {NLU}.
\newblock In \emph{Findings of the Association for Computational Linguistics: {EMNLP}}, pages 4576--4581.

\bibitem[{Huang et~al.(2023)Huang, Shen, Yu, Han, and Liu}]{Huang0Y0L23}
Zhuo Huang, Li~Shen, Jun Yu, Bo~Han, and Tongliang Liu. 2023.
\newblock {FlatMatch}: Bridging labeled data and unlabeled data with cross-sharpness for semi-supervised learning.
\newblock In \emph{Advances in Neural Information Processing Systems 36}.

\bibitem[{Ishimwe et~al.(2021)Ishimwe, Nguyen, and Nguyen}]{IshimweNN21}
Didier Ishimwe, KimHao Nguyen, and ThanhVu Nguyen. 2021.
\newblock Dynaplex: analyzing program complexity using dynamically inferred recurrence relations.
\newblock \emph{Proceedings of the {ACM} on Programming Languages}, 5({OOPSLA}):1--23.

\bibitem[{Kim et~al.(2022)Kim, Woo, Oh, Cha, and Han}]{KimWOCH22}
Hazel~H. Kim, Daecheol Woo, Seong~Joon Oh, Jeong{-}Won Cha, and Yo{-}Sub Han. 2022.
\newblock {ALP:} data augmentation using lexicalized pcfgs for few-shot text classification.
\newblock In \emph{Thirty-Sixth {AAAI} Conference on Artificial Intelligence}, pages 10894--10902.

\bibitem[{Lee(2013)}]{Lee2013}
Dong-Hyun Lee. 2013.
\newblock Pseudo-label: The simple and efficient semi-supervised learning method for deep neural networks.
\newblock In \emph{Workshop on challenges in representation learning, ICML}, page 896.

\bibitem[{Moudgalya et~al.(2023)Moudgalya, Ramakrishnan, Chemudupati, and Lu}]{MoudgalyaRCL23}
Kaushik Moudgalya, Ankit Ramakrishnan, Vamsikrishna Chemudupati, and Xing~Han Lu. 2023.
\newblock {TASTY:} {A} transformer based approach to space and time complexity.
\newblock \emph{preprint}, arXiv: 2305.05379.

\bibitem[{Nie et~al.(2024)Nie, Deng, Liu, Wei, Han, and Zheng}]{NieDLWHZ24}
Wenhua Nie, Lin Deng, Chang{-}Bo Liu, Jialing Wei, Ruitong Han, and Haoran Zheng. 2024.
\newblock {STSPL-SSC:} semi-supervised few-shot short text clustering with semantic text similarity optimized pseudo-labels.
\newblock In \emph{Findings of the Association for Computational Linguistics, {ACL}}, pages 12174--12185.

\bibitem[{OpenAI(2023)}]{gpt-report}
OpenAI. 2023.
\newblock {GPT-4} technical report.
\newblock \emph{preprint}, arXiv: 2303.08774.

\bibitem[{Sajjadi et~al.(2016)Sajjadi, Javanmardi, and Tasdizen}]{SajjadiJT2016}
Mehdi Sajjadi, Mehran Javanmardi, and Tolga Tasdizen. 2016.
\newblock Regularization with stochastic transformations and perturbations for deep semi-supervised learning.
\newblock \emph{Advances in Neural Information Processing Systems 29}, 29.

\bibitem[{Sikka et~al.(2020)Sikka, Satya, Kumar, Uppal, Shah, and Zimmermann}]{SikkaSKUSZ20}
Jagriti Sikka, Kushal Satya, Yaman Kumar, Shagun Uppal, Rajiv~Ratn Shah, and Roger Zimmermann. 2020.
\newblock Learning based methods for code runtime complexity prediction.
\newblock In \emph{Advances in Information Retrieval - 42nd European Conference on {IR} Research, {ECIR} Proceedings, Part {I}}, volume 12035, pages 313--325.

\bibitem[{Sohn et~al.(2020)Sohn, Berthelot, Carlini, Zhang, Zhang, Raffel, Cubuk, Kurakin, and Li}]{SohnBCZZ2020}
Kihyuk Sohn, David Berthelot, Nicholas Carlini, Zizhao Zhang, Han Zhang, Colin~A Raffel, Ekin~Dogus Cubuk, Alexey Kurakin, and Chun-Liang Li. 2020.
\newblock Fixmatch: Simplifying semi-supervised learning with consistency and confidence.
\newblock \emph{Advances in Neural Information Processing Systems 33}, 33:596--608.

\bibitem[{Wang et~al.(2023{\natexlab{a}})Wang, Chen, Heng, Hou, Fan, Wu, Wang, Savvides, Shinozaki, Raj, Schiele, and Xie}]{WangCHH2022}
Yidong Wang, Hao Chen, Qiang Heng, Wenxin Hou, Yue Fan, Zhen Wu, Jindong Wang, Marios Savvides, Takahiro Shinozaki, Bhiksha Raj, Bernt Schiele, and Xing Xie. 2023{\natexlab{a}}.
\newblock Freematch: Self-adaptive thresholding for semi-supervised learning.
\newblock In \emph{The Eleventh International Conference on Learning Representations, {ICLR}}.

\bibitem[{Wang et~al.(2023{\natexlab{b}})Wang, Le, Gotmare, Bui, Li, and Hoi}]{WangLGB0H23}
Yue Wang, Hung Le, Akhilesh Gotmare, Nghi D.~Q. Bui, Junnan Li, and Steven C.~H. Hoi. 2023{\natexlab{b}}.
\newblock Code{T}5+: Open code large language models for code understanding and generation.
\newblock In \emph{Proceedings of the 2023 Conference on Empirical Methods in Natural Language Processing, {EMNLP}}, pages 1069--1088.

\bibitem[{Wei and Zou(2019)}]{WeiZ19}
Jason~W. Wei and Kai Zou. 2019.
\newblock {EDA:} easy data augmentation techniques for boosting performance on text classification tasks.
\newblock In \emph{Proceedings of the 2019 Conference on Empirical Methods in Natural Language Processing and the 9th International Joint Conference on Natural Language Processing, {EMNLP-IJCNLP}}, pages 6381--6387.

\bibitem[{Xie et~al.(2020)Xie, Dai, Hovy, Luong, and Le}]{XieDHLL2020}
Qizhe Xie, Zihang Dai, Eduard Hovy, Thang Luong, and Quoc Le. 2020.
\newblock Unsupervised data augmentation for consistency training.
\newblock \emph{Advances in Neural Information Processing Systems 33}, 33:6256--6268.

\bibitem[{Zhang et~al.(2021{\natexlab{a}})Zhang, Wang, Hou, Wu, Wang, Okumura, and Shinozaki}]{ZhangWHW2021}
Bowen Zhang, Yidong Wang, Wenxin Hou, Hao Wu, Jindong Wang, Manabu Okumura, and Takahiro Shinozaki. 2021{\natexlab{a}}.
\newblock Flexmatch: Boosting semi-supervised learning with curriculum pseudo labeling.
\newblock \emph{Advances in Neural Information Processing Systems 34}, 34:18408--18419.

\bibitem[{Zhang et~al.(2023)Zhang, Wu, and Lu}]{ZhangWL23}
En~Zhang, Fan Wu, and Xuesong Lu. 2023.
\newblock An intelligent scheduling system for large-scale online judging.
\newblock In \emph{Computer Science and Education. Computer Science and Technology - 18th International Conference, {ICCSE}, Proceedings, Part {I}}, volume 2023 of \emph{Communications in Computer and Information Science}, pages 265--279.

\bibitem[{Zhang et~al.(2021{\natexlab{b}})Zhang, Bui, Yoon, Chen, Liu, Xia, Tran, Chang, and Yu}]{ZhangB0CLXTCY21}
Jian{-}Guo Zhang, Trung Bui, Seunghyun Yoon, Xiang Chen, Zhiwei Liu, Congying Xia, Quan~Hung Tran, Walter Chang, and Philip~S. Yu. 2021{\natexlab{b}}.
\newblock Few-shot intent detection via contrastive pre-training and fine-tuning.
\newblock In \emph{Proceedings of the 2021 Conference on Empirical Methods in Natural Language Processing, {EMNLP}}, pages 1906--1912.

\bibitem[{Zhang et~al.(2022)Zhang, Wang, Zhang, Li, and Jin}]{ZhangWZ0J22}
Kechi Zhang, Wenhan Wang, Huangzhao Zhang, Ge~Li, and Zhi Jin. 2022.
\newblock Learning to represent programs with heterogeneous graphs.
\newblock In \emph{Proceedings of the 30th {IEEE/ACM} International Conference on Program Comprehension, {ICPC}}, pages 378--389.

\bibitem[{Zou and Caragea(2023)}]{ZouC23}
Henry~Peng Zou and Cornelia Caragea. 2023.
\newblock {JointMatch}: {A} unified approach for diverse and collaborative pseudo-labeling to semi-supervised text classification.
\newblock In \emph{Proceedings of the 2023 Conference on Empirical Methods in Natural Language Processing, {EMNLP}}, pages 7290--7301.

\bibitem[{Zou et~al.(2023)Zou, Caragea, Zhou, and Caragea}]{ZouCZC23}
Henry~Peng Zou, Cornelia Caragea, Yue Zhou, and Doina Caragea. 2023.
\newblock Semi-supervised few-shot learning for fine-grained disaster tweet classification.
\newblock In \emph{Proceedings of the 20th International ISCRAM Conference}.

\end{thebibliography}

\onecolumn

\appendix
\section{\texttt{TCProF} Full Result}\label{app:full-result}
\texttt{TCProF} consists of three core parts.
The first is a symbolic module~$\mathbb{Sym}$ for pseudo-labeling.
The second is a data augmentation module~$\mathbb{AUG}$ which employs
Back-Translation~(BT), Loop-Conversion~(LC), and both~(BT+LC).
Finally, the third one is a co-training module.
Our hypothesis is that the more components implemented, the better the performance
and eventually, \texttt{TCProF}, with all modules implemented, performs the best.
The result of ST(CodeBERT) and ST(GraphCodeBERT) were omitted in
Table~\ref{tbl:baseline-comparison-acc-part},
and the result of $\mathbb{AUG}_{BT}$ and$\mathbb{AUG}_{LC}$
were omitted in Table~\ref{tbl:ablation-studies}.
We present the full experimental results regarding all the three components of \texttt{TCProF} in
Tables~\ref{tbl:baseline-comparison-acc-full} and~\ref{tbl:baseline-comparison-f1},
each for the accuracy and F1 performance.

\begin{table*}[ht]
\centering
\resizebox{\textwidth}{!}{%
\begin{tabular}{lcccccc}
\hline
               &  \multicolumn{2}{c}{CodeComplex~(Java)}  &  \multicolumn{2}{c}{CodeComplex~(Python)}  &        \multicolumn{2}{c}{CorCoD}           \\
               &          5          &         10         &          5           &         10          &           5           &         10          \\
\hline
\multicolumn{7}{l}{SSL Baselines} \\
\hline
ST(CodeBERT)                            &22.82\tiny$\pm$ 0.00  &26.47\tiny$\pm$ 6.33 &16.33\tiny$\pm$  0.24  &15.51\tiny$\pm$  1.18 &35.79\tiny$\pm$  0.00   &35.79\tiny$\pm$  0.00 \\
ST(GraphCodeBERT)                       &22.89\tiny$\pm$ 0.13  &20.51\tiny$\pm$ 5.67 &16.19\tiny$\pm$  0.00  &23.57\tiny$\pm$ 12.78 &36.14\tiny$\pm$  0.61   &35.79\tiny$\pm$  0.00 \\
ST(CodeT5+)                             &18.87\tiny$\pm$ 3.21  &31.69\tiny$\pm$14.00 &28.76\tiny$\pm$ 17.47  &38.32\tiny$\pm$ 18.82 &35.79\tiny$\pm$  6.90   &45.26\tiny$\pm$  6.32 \\
ST(UniXcoder)                           &15.44\tiny$\pm$ 6.30  &31.92\tiny$\pm$ 6.14 &26.02\tiny$\pm$ 11.72  &40.98\tiny$\pm$  1.14 &37.89\tiny$\pm$  6.57   &45.61\tiny$\pm$ 11.59 \\
JointMatch~(CodeT5+)                    &14.62\tiny$\pm$ 4.26  &24.68\tiny$\pm$ 2.60 &20.97\tiny$\pm$  3.90  &21.04\tiny$\pm$  4.33 &36.49\tiny$\pm$  4.98   &42.11\tiny$\pm$  6.57 \\
JointMatch~(UniXcoder)                  & 9.62\tiny$\pm$ 5.23  &19.39\tiny$\pm$ 3.86 &14.68\tiny$\pm$ 11.78  &20.76\tiny$\pm$ 10.46 &35.44\tiny$\pm$  6.85   &48.42\tiny$\pm$  8.22 \\
\hline
\hline
\multicolumn{7}{c}{CodeT5+} \\
\hline
\hspace{4pt}+ \hspace{2pt}$\mathbb{AUG}_{BT}$                         &20.13\tiny$\pm$ 3.31  & 26.77\tiny$\pm$ 5.38&  39.34\tiny$\pm$19.17 &45.36\tiny$\pm$11.58  &34.73\tiny$\pm$ 7.59    &38.94\tiny$\pm$ 2.79 \\
\hspace{4pt}+ \hspace{2pt}$\mathbb{AUG}_{LC}$                         &26.47\tiny$\pm$ 5.49  & 29.53\tiny$\pm$ 5.83&  32.92\tiny$\pm$14.47 &53.89\tiny$\pm$16.03  &35.78\tiny$\pm$ 5.47    &42.45\tiny$\pm$ 7.97 \\
\hspace{4pt}+ \hspace{2pt}$\mathbb{AUG}_{BT+LC}$                      &26.85\tiny$\pm$ 2.4   &35.86\tiny$\pm$ 4.53 &  33.94\tiny$\pm$ 3.82 &57.44\tiny$\pm$ 1.17  &36.14\tiny$\pm$ 4.98    &43.85\tiny$\pm$ 6.34 \\
\hspace{4pt}+ \hspace{2pt}$\mathbb{Sym}$                              &21.92\tiny$\pm$ 2.54  &35.50\tiny$\pm$11.18 &  32.10\tiny$\pm$ 6.78 &46.38\tiny$\pm$ 6.58  &36.84\tiny$\pm$ 1.72    &42.81\tiny$\pm$ 2.76 \\
\hspace{4pt}+ \hspace{2pt}$\mathbb{Sym}$ + $\mathbb{AUG}_{BT+LC}$     &37.88\tiny$\pm$ 1.65  &38.26\tiny$\pm$ 2.40 &  41.80\tiny$\pm$ 1.49 &58.88\tiny$\pm$12.37  &37.89\tiny$\pm$ 2.58    &44.21\tiny$\pm$ 2.27 \\
\hspace{4pt}\texttt{TCProF}(CodeT5+)                                  &38.63\tiny$\pm$ 1.32  &41.98\tiny$\pm$ 2.92 &  44.74\tiny$\pm$ 3.26 &59.29\tiny$\pm$ 3.71  &50.53\tiny$\pm$ 0.86    &51.93\tiny$\pm$ 3.88 \\
\hline
\hline
\multicolumn{7}{c}{UnixCoder} \\
\hline
\hspace{4pt}+ \hspace{2pt}$\mathbb{AUG}_{BT}$                         &29.75\tiny$\pm$ 3.17  &33.40\tiny$\pm$ 14.43&50.31\tiny$\pm$19.27   &52.93\tiny$\pm$10.66  &33.15\tiny$\pm$ 3.72    &50.17\tiny$\pm$ 9.43 \\
\hspace{4pt}+ \hspace{2pt}$\mathbb{AUG}_{LC}$                         &30.72\tiny$\pm$ 8.40  &33.03\tiny$\pm$ 6.55 &44.53\tiny$\pm$13.40   &53.55\tiny$\pm$ 4.10  &38.94\tiny$\pm$ 3.16    &47.01\tiny$\pm$ 4.38 \\
\hspace{4pt}+ \hspace{2pt}$\mathbb{AUG}_{BT+LC}$                      &34.22\tiny$\pm$12.32  &39.00\tiny$\pm$ 6.64 &51.63\tiny$\pm$12.42   &63.04\tiny$\pm$ 8.75  &44.21\tiny$\pm$ 3.16    &51.57\tiny$\pm$ 5.57 \\
\hspace{4pt}+ \hspace{2pt}$\mathbb{Sym}$                              &41.76\tiny$\pm$ 2.62  &45.34\tiny$\pm$ 2.91 &39.82\tiny$\pm$ 5.13   &51.02\tiny$\pm$ 2.14  &50.88\tiny$\pm$ 8.48    &54.04\tiny$\pm$ 6.56 \\
\hspace{4pt}+ \hspace{2pt}$\mathbb{Sym}$ + $\mathbb{AUG}_{BT+LC}$     &43.70\tiny$\pm$ 2.00  &45.49\tiny$\pm$ 1.10 &54.03\tiny$\pm$ 5.52   &67.55\tiny$\pm$ 1.06  &50.18\tiny$\pm$ 4.24    &58.60\tiny$\pm$ 4.73 \\
\hspace{4pt}\texttt{TCProF}(UniXcoder)                                &\textbf{52.50}\tiny$\pm$ 1.56  &\textbf{53.85}\tiny$\pm$ 3.63 &\textbf{54.64}\tiny$\pm$ 3.77   &\textbf{70.29}\tiny$\pm$ 2.06  &\textbf{55.44}\tiny$\pm$ 0.99    &\textbf{63.16}\tiny$\pm$ 2.27 \\
\hline
\end{tabular}
}
\caption{Accuracy performance comparisons. The scores are averaged from three runs with different seeds.}
\label{tbl:baseline-comparison-acc-full}
\end{table*}

\begin{figure*}[ht]
    \centering
    \begin{subfigure}[b]{.32\textwidth}
    \includegraphics[width=\textwidth]{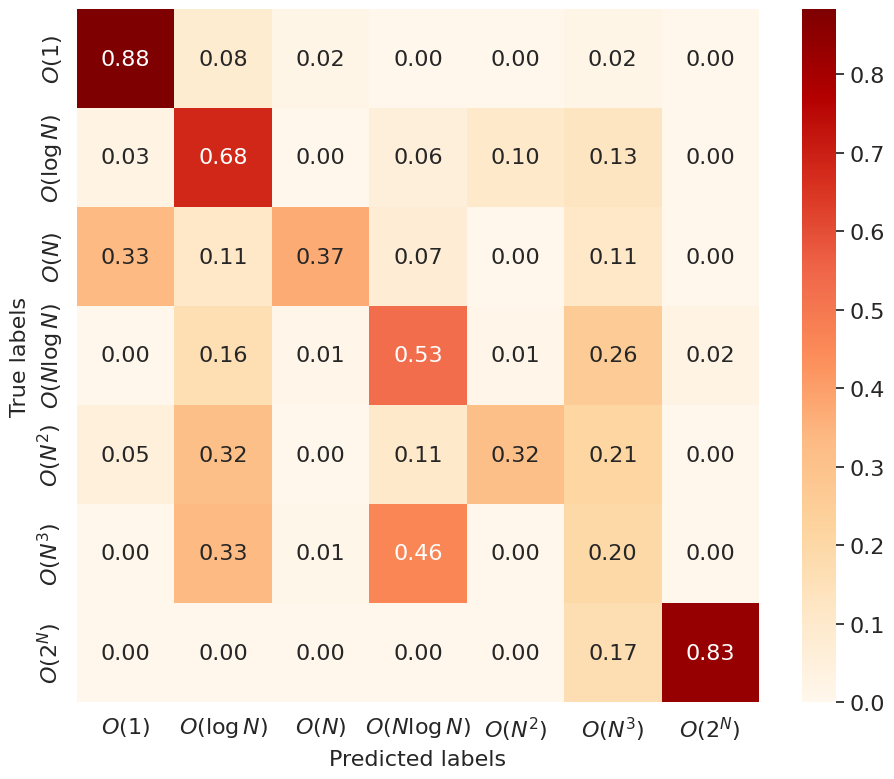}
    \caption{CodeComplex~(Java).}
    \label{fig:unixcoder-java-confusion}
    \end{subfigure}
    \hfill
    \begin{subfigure}[b]{.32\textwidth}
    \includegraphics[width=\textwidth]{Figures/python_cross_unixcoder_confusion.png}
    \caption{CodeComplex~(Python).}
    \label{fig:unixcoder-python-confusion}
    \end{subfigure}
    \hfill
    \begin{subfigure}[b]{.32\textwidth}
    \includegraphics[width=\textwidth]{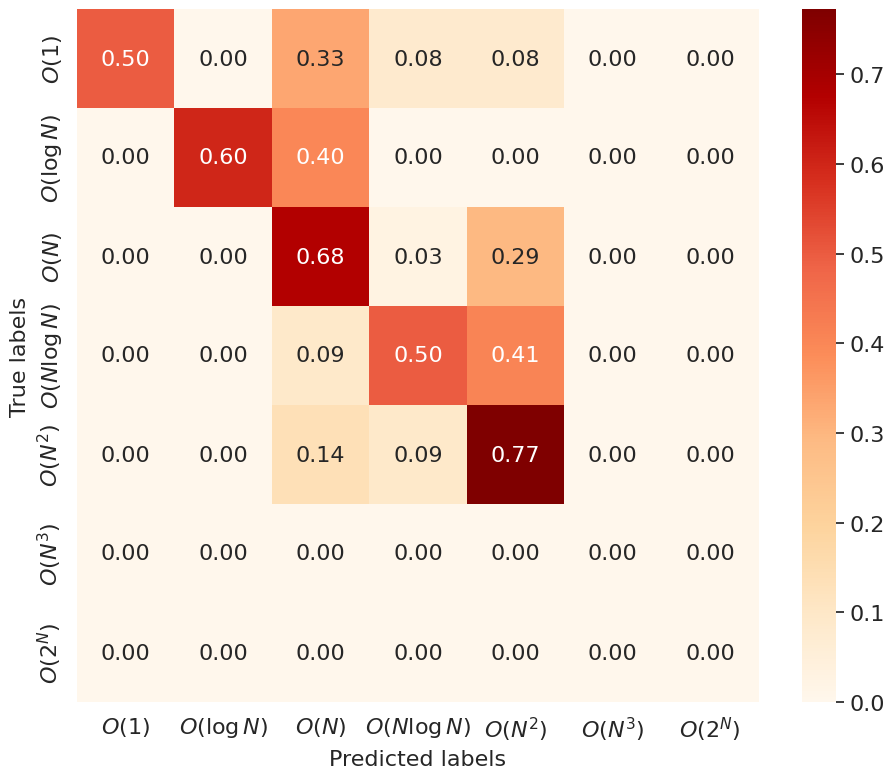}
    \caption{CorCoD.}
    \label{fig:unixcoder-corcod-confusion}
    \end{subfigure}
    \caption{Confusion matrices of 10-shot \texttt{TCProF}(UniXcoder) performance.
    }
\end{figure*}

Both results are consistent with our hypothesis.
Comparing the performance of $\mathbb{AUG}_{BT}$ and $\mathbb{AUG}_{LC}$ with $\mathbb{AUG}_{BT+LC}$,
the tendency of performance scores demonstrates that using both augmentation methods is superior
to using the single method.
Additional implementation of $\mathbb{Sym}$ also improves the scores.
It is notable that some performance scores of $\mathbb{AUG}_{BT+LC}$ with high standard deviation scores
show lesser standard deviation scores when implemented with $\mathbb{Sym}$.
This indicates that the usage of symbolic pseudo-labels and augmentation complements each other and enhances the model performance.
Finally, with all modules and co-training strategy implemented, \texttt{TCProF}, performs the best.

\begin{table*}[ht]
\centering
\resizebox{\columnwidth}{!}{%
\begin{tabular}{lllllll}
\hline
               &  \multicolumn{2}{c}{CodeComplex~(Java)}  &  \multicolumn{2}{c}{CodeComplex~(Python)}  &        \multicolumn{2}{c}{CorCoD}           \\
               &\multicolumn{1}{c}{5}&\multicolumn{1}{c}{10}&\multicolumn{1}{c}{5}&\multicolumn{1}{c}{10}&\multicolumn{1}{c}{5}&\multicolumn{1}{c}{10}          \\
\hline
\multicolumn{7}{l}{SSL Baselines} \\
\hline
ST(CodeBERT)                            & 5.31\tiny$\pm$ 0.00  & 8.89\tiny$\pm$ 6.20 & 4.27\tiny$\pm$  0.51  & 4.50\tiny$\pm$  0.90 &10.54\tiny$\pm$  0.00   &10.54\tiny$\pm$  0.00 \\
ST(GraphCodeBERT)                       & 5.52\tiny$\pm$ 0.29  &12.33\tiny$\pm$11.97 & 3.98\tiny$\pm$  0.00  & 9.89\tiny$\pm$ 10.24 &12.82\tiny$\pm$  3.94   &10.54\tiny$\pm$  0.00 \\
ST(CodeT5+)                             &18.94\tiny$\pm$ 3.69  &29.20\tiny$\pm$10.37 &23.74\tiny$\pm$ 12.26  &28.48\tiny$\pm$ 11.05 &26.87\tiny$\pm$ 17.65   &45.17\tiny$\pm$  7.10 \\
ST(UniXcoder)                           &13.52\tiny$\pm$ 6.65  &34.91\tiny$\pm$ 5.73 &21.61\tiny$\pm$  9.43  &34.57\tiny$\pm$  0.80 &38.92\tiny$\pm$  7.24   &51.16\tiny$\pm$  8.61 \\
JointMatch~(CodeT5+)                    & 7.43\tiny$\pm$ 1.26  &25.24\tiny$\pm$ 1.51 &13.61\tiny$\pm$  6.05  &14.46\tiny$\pm$  3.18 &30.31\tiny$\pm$  8.01   &45.86\tiny$\pm$  7.78 \\
JointMatch~(UniXcoder)                  & 9.70\tiny$\pm$ 5.41  &10.38\tiny$\pm$ 5.20 & 7.87\tiny$\pm$  8.82  &13.02\tiny$\pm$  8.85 &18.98\tiny$\pm$  7.43   &39.44\tiny$\pm$ 11.93 \\
\hline
\hline
\multicolumn{7}{c}{CodeT5+} \\
\hline
\hspace{4pt}+ \hspace{2pt}$\mathbb{AUG}_{BT}$                         &22.90\tiny$\pm$ 1.60  &27.31\tiny$\pm$ 3.13 &31.94\tiny$\pm$ 9.17   &35.83\tiny$\pm$ 4.52  &35.80\tiny$\pm$10.98    &43.85\tiny$\pm$ 2.99 \\
\hspace{4pt}+ \hspace{2pt}$\mathbb{AUG}_{LC}$                         &25.33\tiny$\pm$ 3.56  &26.42\tiny$\pm$ 4.02 &26.60\tiny$\pm$ 5.77   &42.50\tiny$\pm$ 8.51  &39.36\tiny$\pm$ 6.15    &46.76\tiny$\pm$ 6.72 \\
\hspace{4pt}+ \hspace{2pt}$\mathbb{AUG}_{BT+LC}$                      &26.23\tiny$\pm$ 2.07  &30.69\tiny$\pm$ 8.08 &31.12\tiny$\pm$ 5.61   &42.38\tiny$\pm$ 1.11  &41.78\tiny$\pm$ 5.12    &47.98\tiny$\pm$ 4.70 \\
\hspace{4pt}+ \hspace{2pt}$\mathbb{Sym}$                              &19.54\tiny$\pm$ 1.76  &30.52\tiny$\pm$10.61 &23.18\tiny$\pm$ 2.75   &34.69\tiny$\pm$ 3.51  &39.86\tiny$\pm$ 1.77    &46.66\tiny$\pm$ 4.20 \\
\hspace{4pt}+ \hspace{2pt}$\mathbb{Sym}$ + $\mathbb{AUG}_{BT+LC}$     &29.11\tiny$\pm$ 1.64  &32.72\tiny$\pm$ 1.48 &31.99\tiny$\pm$ 1.64   &42.95\tiny$\pm$10.01  &42.36\tiny$\pm$ 8.08    &48.09\tiny$\pm$ 1.78 \\
\hspace{4pt}\texttt{TCProF}(CodeT5+)                                  &28.58\tiny$\pm$ 0.37  &33.60\tiny$\pm$ 0.98 &34.49\tiny$\pm$ 4.55   &44.03\tiny$\pm$ 2.74  &45.13\tiny$\pm$ 5.27    &49.60\tiny$\pm$ 8.12 \\
\hline
\hline
\multicolumn{7}{c}{UnixCoder} \\
\hline
\hspace{4pt}+ \hspace{2pt}$\mathbb{AUG}_{BT}$                         &29.18\tiny$\pm$ 3.95  &33.52\tiny$\pm$ 10.03 &31.43\tiny$\pm$ 5.98   &39.24\tiny$\pm$ 3.60  &38.82\tiny$\pm$ 9.97    &52.37\tiny$\pm$ 8.00  \\
\hspace{4pt}+ \hspace{2pt}$\mathbb{AUG}_{LC}$                         &27.24\tiny$\pm$2.77   &36.26\tiny$\pm$ 16.28 &29.48\tiny$\pm$ 3.87   &34.82\tiny$\pm$ 4.26  &40.33\tiny$\pm$ 8.51    &48.20\tiny$\pm$13.78  \\
\hspace{4pt}+ \hspace{2pt}$\mathbb{AUG}_{BT+LC}$                      &31.02\tiny$\pm$14.22  &34.21\tiny$\pm$ 10.92 &33.53\tiny$\pm$ 5.73   &45.95\tiny$\pm$ 8.69  &43.42\tiny$\pm$ 1.29    &52.57\tiny$\pm$ 5.91  \\
\hspace{4pt}+ \hspace{2pt}$\mathbb{Sym}$                              &31.88\tiny$\pm$ 1.28  &44.74\tiny$\pm$ 2.10  &24.50\tiny$\pm$ 3.48   &40.77\tiny$\pm$ 2.56  &48.77\tiny$\pm$ 6.32    &54.12\tiny$\pm$ 6.47  \\
\hspace{4pt}+ \hspace{2pt}$\mathbb{Sym}$ + $\mathbb{AUG}_{BT+LC}$     &39.66\tiny$\pm$ 0.44  &45.31\tiny$\pm$ 4.48  &37.64\tiny$\pm$ 4.90   &51.58\tiny$\pm$ 1.61  &47.45\tiny$\pm$ 4.93    &56.87\tiny$\pm$10.95  \\
\hspace{4pt}\texttt{TCProF}(UniXcoder)                                &\textbf{42.89}\tiny$\pm$ 3.09  &\textbf{49.45}\tiny$\pm$ 0.97 &\textbf{38.15}\tiny$\pm$ 4.09   &\textbf{53.17}\tiny$\pm$ 2.13  &\textbf{56.56}\tiny$\pm$ 3.52    &\textbf{63.57}\tiny$\pm$ 3.09  \\
\hline
\end{tabular}
}
\caption{F1-score performance comparisons. The scores are averaged from three runs with different seeds.}
\label{tbl:baseline-comparison-f1}
\end{table*}

It is remarkable that our hypothesis holds on both the accuracy and F1 performance of \texttt{TCProF}.
Intriguing facts are that while the SSL baselines tend to have higher standard deviation scores for F1 performance than accuracy,
\texttt{TCProF} generally shows less standard deviation scores for F1 performance.
This is because F1-scores evaluate how well the model performs for `all' classes while accuracy does not consider
whether the model is biased on specific classes.
We provide confusion matrices of 10-shot \texttt{TCProF}(UniXcoder) for CodeComplex~(Java), CodeComplex~(Python),
and CorCoD in Figures~\ref{fig:unixcoder-java-confusion},~\ref{fig:unixcoder-python-confusion}, and~\ref{fig:unixcoder-corcod-confusion}, respectively.
We can see that \texttt{TCProF} produces unbiased performance compared to those seen in Figures~\ref{fig:unixcoder-confusion} and~\ref{fig:unixcoder-jointmath-confusion}.

\section{Baseline Selection for Self-Training}\label{app:baseline-selection}
From the four well-known code language models, CodeBERT, GraphCodeBERT, UniXcoder, and CodeT5+,
we experimented with which baseline models would fit for semi-supervised learning~(SSL) in low-resource settings.
Generally, UniXcoder and CodeT5+ are better than CodeBERT and GraphCodeBERT.
We can also see the same tendency comparing the performance of each model trained with the full data in Table~\ref{tbl:baseline-augment-full}.
Additionally analyzing for low-resource settings, we selected two best-performing models, UniXcoder and CodeT5+ to implement \texttt{TCProF}.

CodeBERT and GraphCodeBERT have satisfactory accuracy for 5-shot CodeComplex~(Java) and CorCoD
compared to the other two models as shown in Table~\ref{tbl:baseline-comparison-acc-full}.
However, Table~\ref{tbl:baseline-comparison-f1} reveals that CodeBERT and GraphCodeBERT
perform significantly lower than that of the other two models.
The large performance gap of CodeBERT and GraphCodeBERT between Tables~\ref{tbl:baseline-comparison-acc-full}
and~\ref{tbl:baseline-comparison-f1} and the low F1 performance of these two models
indicate that they are unsuitable to implement our methodology.

\begin{figure}[ht]
    \centering
    \begin{subfigure}[b]{.44\textwidth}
    \includegraphics[width=\textwidth]{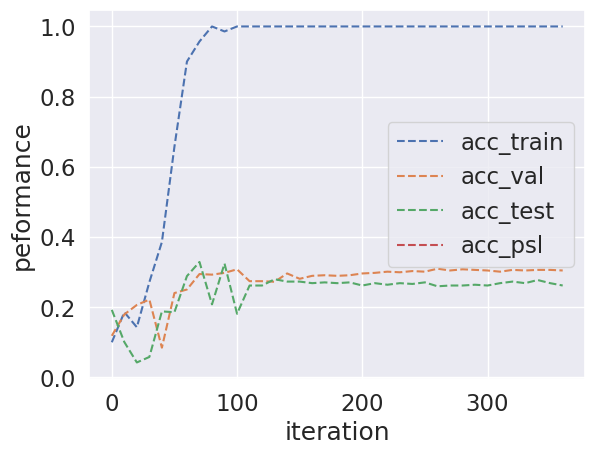}
    \caption{Learning curves and accuracy shifts in CodeT5+ self-training }
    \label{fig:codet5p-acc}
    \end{subfigure}
    \hfill
    \begin{subfigure}[b]{.44\textwidth}
    \includegraphics[width=\textwidth]{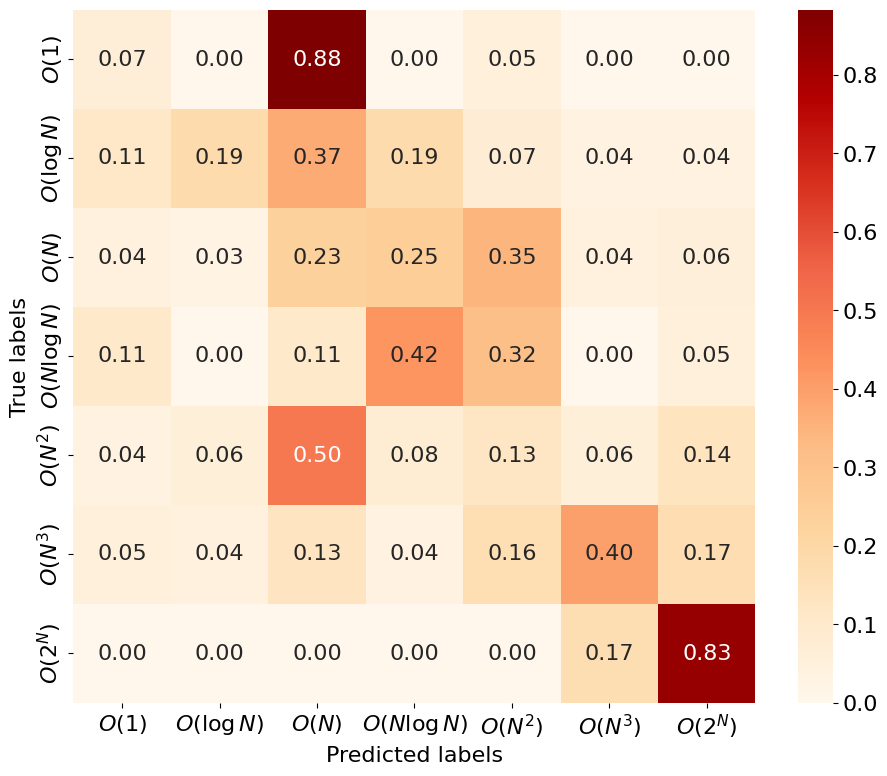}
    \caption{Confusion matrix of CodeT5+ self-training}
    \label{fig:codet5p-confusion}
    \end{subfigure}
    \caption{CodeT5+ self-learning performance visualization for CodeComplex (Java) 10-shot.}
\end{figure}

\begin{figure}[h!]
    \centering
    \begin{subfigure}[b]{.44\textwidth}
    \includegraphics[width=\textwidth]{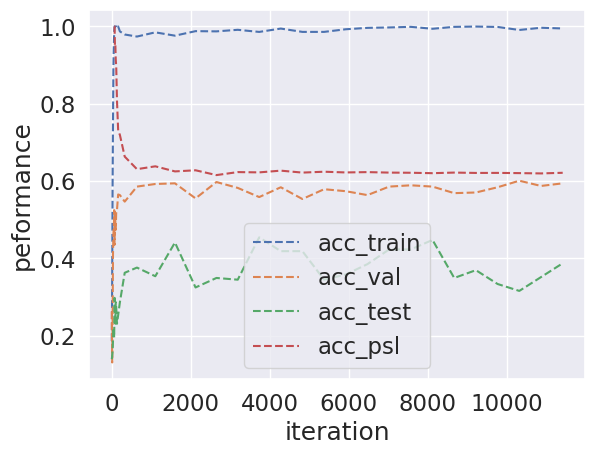}
    \caption{Learning curves and accuracy shifts in UniXcoder self-training }
    \label{fig:unixcoder-acc}
    \end{subfigure}
    \hfill
    \begin{subfigure}[b]{.44\textwidth}
    \includegraphics[width=\textwidth]{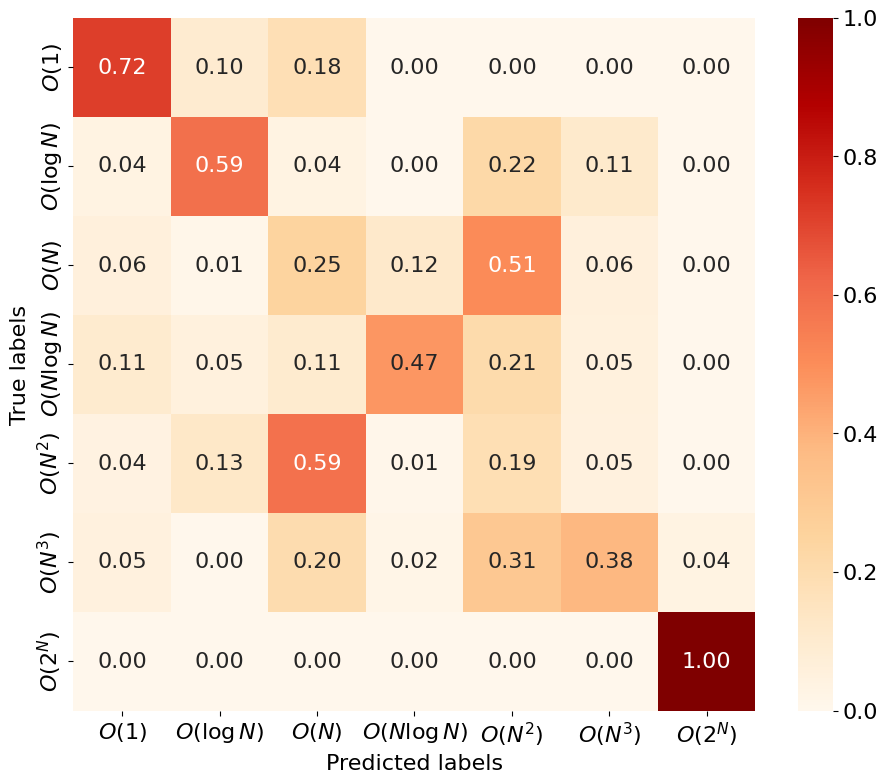}
    \caption{Confusion matrix of UniXcoder self-training}
    \label{fig:unixcoder-confusion}
    \end{subfigure}
    \caption{UniXcoder self-learning performance visualization for CodeComplex~(Java) 10-shot.}
\end{figure}

The relatively high performance of UniXcoder and CodeT5+ in Tables~\ref{tbl:baseline-comparison-acc-full} and~\ref{tbl:baseline-comparison-f1}
empirically prove that they are suitable for low-resource settings.
From confusion matrices of the two models in Figures~\ref{fig:codet5p-confusion} and~\ref{fig:unixcoder-confusion},
we can also easily see that UniXcoder and CodeT5+ do not overfit to a certain class.
Furthermore, we depict the learning processes of these two models in Figures~\ref{fig:codet5p-acc} and~\ref{fig:unixcoder-acc}.
In the figures, \texttt{acc\_train}, \texttt{acc\_val}, \texttt{acc\_test}, and \texttt{acc\_psl} indicate
the accuracy of the train, valid, test, and pseudo-labeled datasets, respectively.
These figures demonstrate that the converging patterns of \texttt{acc\_train}, \texttt{acc\_val}, \texttt{acc\_test}
are similar, indicating that the two models are both suitable for training in low-resource settings.
From these two models, we also analyze which model is more suitable for
low-resource settings, and eventually \texttt{TCProF}.

In Figure~\ref{fig:codet5p-acc}, \texttt{acc\_psl} is not shown in the graph,
meaning that CodeT5+ does not produce pseudo-labels in the self-training at all.
This is because pseudo-labels are produced based on the confidence score and
the confidence score of CodeT5+ does not match the pre-defined threshold.
It is quite surprising as CodeT5+ has high \texttt{acc\_train}.
We have run an experiment of training CodeT5+ continuously to inspect when CodeT5+ generates pseudo-labels and
the model produced biased pseudo-labels for only a few classes for CodeComplex~(Python).
Thus, independent to the performance, CodeT5+ is not suitable for pseudo-labeling when implemented with only self-training.

UniXcoder, on the other hand, surely generates pseudo-labels referring to Figure~\ref{fig:unixcoder-acc}.
For all datasets, UniXcoder generates pseudo-labels for all 7 classes, indicating that
the model shows consistent performance without being biased toward a specific class.
Figure~\ref{fig:unixcoder-confusion} also confirms that UniXcoder shows relatively well-distributed performance for each class in test data.
This confirms that UniXcoder is more suitable that CodeT5+ in a low-resource setting.

Referring to Table~\ref{tbl:baseline-comparison-acc-part} in Section~\ref{ssec:baseline-comparison},
we can see that our proposed framework~\texttt{TCProF} effectively improves the performance of both UniXcoder and CodeT5+
but we can also notice that UniXcoder is more effective for \texttt{TCProF} and eventually, low-resource settings.

\section{Comparison to Baselines Fine-Tuned with Full Dataset}\label{app:full-baseline-finetune}
\begin{figure*}[ht]
    \centering
    \begin{subfigure}[b]{.49\textwidth}
    \includegraphics[width=\textwidth]{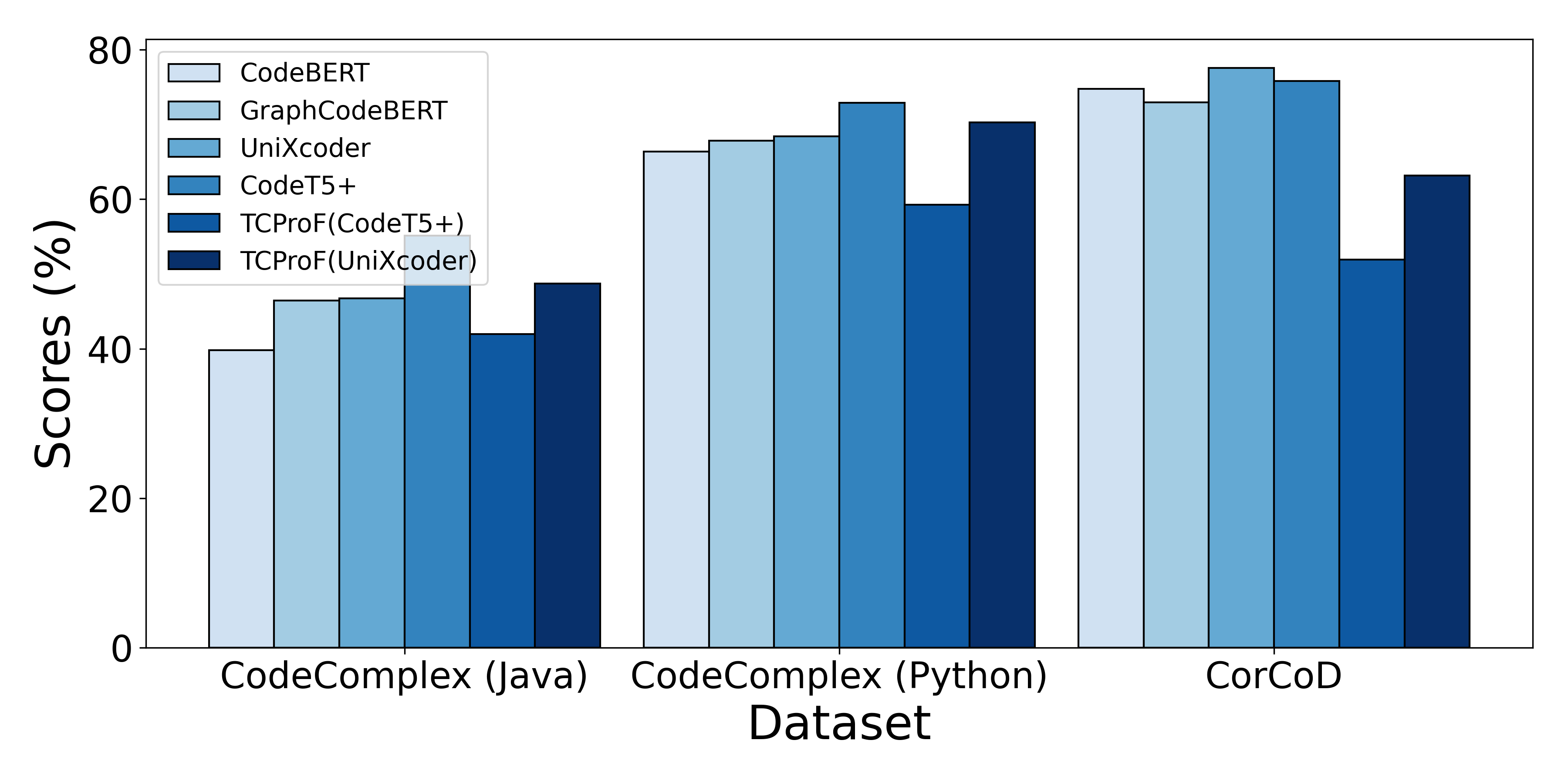}
    \caption{Accuracy comparison.}
    \label{fig:full-acc-graph}
    \end{subfigure}
    \hfill
    \begin{subfigure}[b]{.49\textwidth}
    \includegraphics[width=\textwidth]{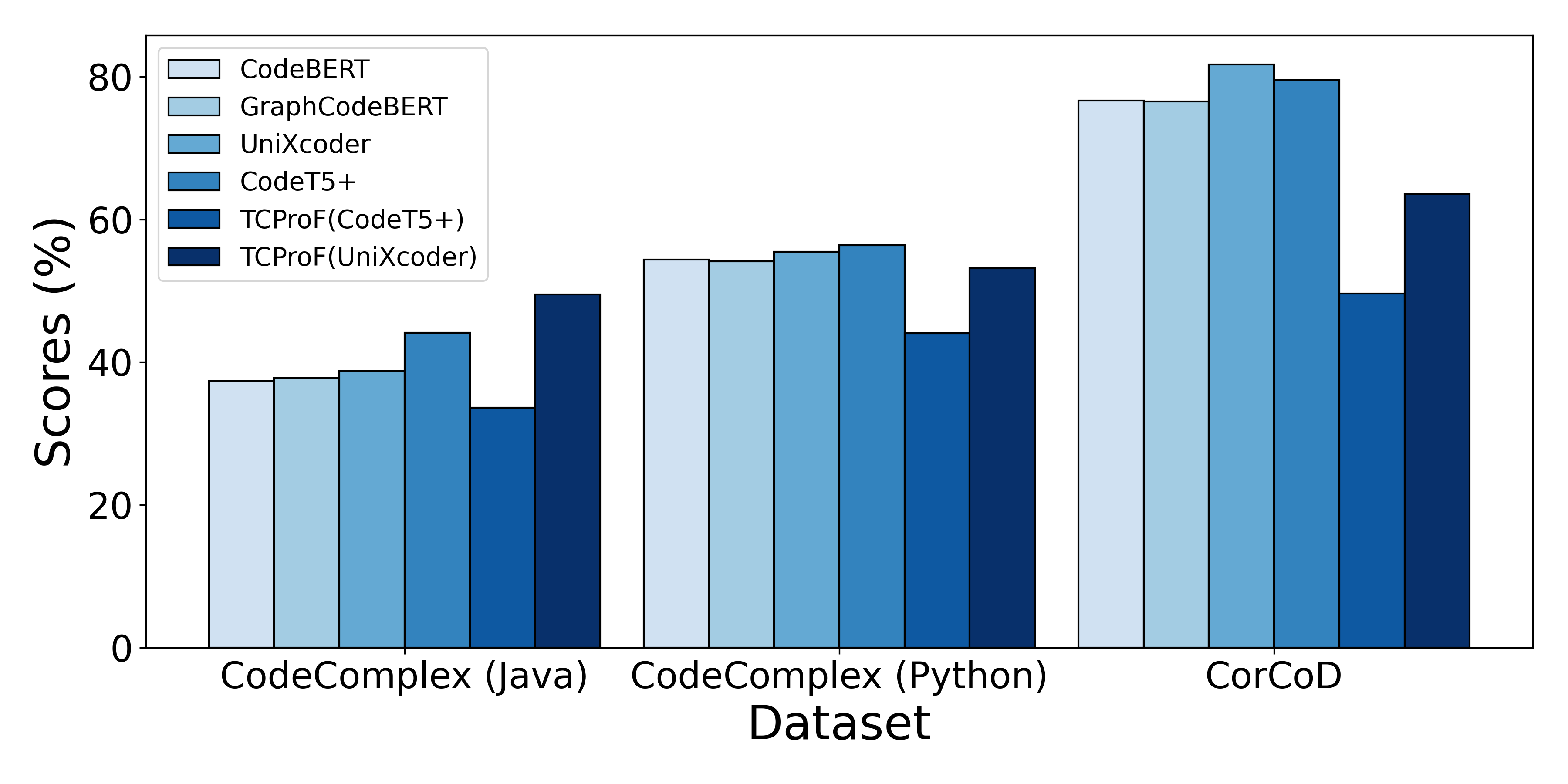}
    \caption{Fl-score comparison.}
    \label{fig:full-f1-graph}
    \end{subfigure}
    \caption{Visualized performance of baselines trained with full train dataset and \texttt{TCProF} 10-shot.
    }
\end{figure*}

We have provided a comparative analysis of \texttt{TCProF} and baselines fine-tuned with the entire train dataset in Section~\ref{ssec:full-comparison}.
Detailed results for accuracy and F1-scores are shown in Table~\ref{tbl:baseline-comparison-full} and Figures~\ref{fig:full-acc-graph} and~\ref{fig:full-f1-graph} provide
full results of accuracy and F1 performance.

\begin{table*}[ht]
\centering
\begin{tabular}{lllllll}
\hline
               &  \multicolumn{2}{c}{CodeComplex~(Java)}  &  \multicolumn{2}{c}{CodeComplex~(Python)}  &        \multicolumn{2}{c}{CorCoD}           \\
               &         Acc.         &         F1          &         Acc.          &         F1           &          Acc.          &         F1          \\
\hline
\multicolumn{7}{l}{Baselines trained with the Full train dataset} \\
\hline
CodeBERT        &39.82\tiny$\pm$ 6.90 &37.35\tiny$\pm$ 3.56 &66.39\tiny$\pm$  1.43  &54.34\tiny$\pm$  0.93 &74.74\tiny$\pm$  1.72   &76.64\tiny$\pm$  3.07 \\
GraphCodeBERT   &46.46\tiny$\pm$ 2.18 &37.75\tiny$\pm$ 1.18 &67.83\tiny$\pm$  8.36  &54.13\tiny$\pm$  7.25 &72.98\tiny$\pm$  1.79   &76.48\tiny$\pm$  2.26 \\
UniXcoder       &46.76\tiny$\pm$ 1.91 &38.76\tiny$\pm$ 0.39 &68.44\tiny$\pm$  4.66  &55.45\tiny$\pm$  3.33 &\textbf{77.54}\tiny$\pm$  1.31   &\textbf{81.69}\tiny$\pm$  1.43 \\
CodeT5+         &\textbf{55.11}\tiny$\pm$ 2.18 &44.14\tiny$\pm$ 3.54 &\textbf{72.88}\tiny$\pm$  1.95  &\textbf{56.39}\tiny$\pm$  2.25 &75.79\tiny$\pm$  2.27   &79.51\tiny$\pm$  1.96 \\
\hline
\multicolumn{7}{l}{10-shot performance of \texttt{TCProF}} \\
\hline
\texttt{TCProF}(CodeT5+)         &41.98\tiny$\pm$2.92  &        33.60 \tiny$\pm$0.98  &59.29\tiny$\pm$3.71    &44.03\tiny$\pm$2.74   &  51.93\tiny$\pm$ 3.88  &49.60\tiny$\pm$ 8.12  \\
\texttt{TCProF}(UniXcoder)       &53.85\tiny$\pm$3.63  &\textbf{49.45}\tiny$\pm$0.97  &70.29\tiny$\pm$2.06    &53.17\tiny$\pm$2.13   &  63.16\tiny$\pm$ 2.27  &63.57\tiny$\pm$ 3.09  \\
\hline
\end{tabular}
\caption{Performance comparison of baselines trained with the full-train dataset and \texttt{TCProF} 10-shot performance.}
\label{tbl:baseline-comparison-full}
\end{table*}

We can see that \texttt{TCProF} effectively catches up with both the performance of accuracy and F1-scores.
It is remarkable \texttt{TCProF} even exceeds the F1 performance of CodeT5+ for CodeComplex~(Java),
accomplishing the best F1-scores for the dataset.

\section{Baselines Fine-Tuned with the Augmented Full Dataset}
We have implemented $\mathbb{AUG}_{BT+LC}$,
which combines Back-Translation~(BT) and Loop-Conversion~(LC)
to investigate the impact
of data augmentation on the full dataset.
The results in Table~\ref{tbl:baseline-augment-full} demonstrate
the effectiveness of augmentation across the baselines and datasets.
Notably, CodeT5+ shows a significant enhancement, particularly in Python,
where it achieves over 90\% accuracy and 85\% F1-scores.
This improvement is consistent across different models,
indicating that $\mathbb{AUG}_{BT+LC}$ is
beneficial in predicting time complexity more accurately and reliably.

\begin{table*}[ht]
\centering
\begin{tabular}{lcccccc}
\toprule
               &  \multicolumn{2}{c}{CodeComplex~(Java)}  &  \multicolumn{2}{c}{CodeComplex~(Python)}  &        \multicolumn{2}{c}{CorCoD}           \\
               &         Acc.         &         F1          &         Acc.          &         F1           &          Acc.          &         F1          \\
\midrule
\multicolumn{7}{l}{Original} \\
\hline
CodeBERT        &39.82\tiny$\pm$ 6.90 &37.35\tiny$\pm$ 3.56 &66.39\tiny$\pm$  1.43  &54.34\tiny$\pm$  0.93 &74.74\tiny$\pm$  1.72   &76.64\tiny$\pm$  3.07 \\
GraphCodeBERT   &46.46\tiny$\pm$ 2.18 &37.75\tiny$\pm$ 1.18 &67.83\tiny$\pm$  8.36  &54.13\tiny$\pm$  7.25 &72.98\tiny$\pm$  1.79   &76.48\tiny$\pm$  2.26 \\
UniXcoder       &46.76\tiny$\pm$ 1.91 &38.76\tiny$\pm$ 0.39 &68.44\tiny$\pm$  4.66  &55.45\tiny$\pm$  3.33 &77.54\tiny$\pm$  1.31   &81.69\tiny$\pm$  1.43 \\
CodeT5+         &55.11\tiny$\pm$ 2.18 &44.14\tiny$\pm$ 3.54 &72.88\tiny$\pm$  1.95  &56.39\tiny$\pm$  2.25 &75.79\tiny$\pm$  2.27   &79.51\tiny$\pm$  1.96 \\
\hline
\hline
\multicolumn{7}{l}{Augmented} \\
\hline
CodeBERT        &50.41\tiny$\pm$ 2.84 &42.11\tiny$\pm$ 1.99 &81.83\tiny$\pm$  1.79  &69.20\tiny$\pm$  4.07 &78.60\tiny$\pm$  0.99   &80.06\tiny$\pm$  0.69 \\
GraphCodeBERT   &58.69\tiny$\pm$ 2.89 &46.96\tiny$\pm$ 2.25 &85.31\tiny$\pm$  0.59  &77.82\tiny$\pm$  0.62 &78.60\tiny$\pm$  0.99   &81.08\tiny$\pm$  0.80 \\
UniXcoder       &\textbf{63.91}\tiny$\pm$ 1.28 &\textbf{49.07}\tiny$\pm$ 1.18 &86.13\tiny$\pm$  1.58  &77.47\tiny$\pm$  2.80 &78.95\tiny$\pm$  0.00   &82.22\tiny$\pm$  0.50 \\
CodeT5+         &59.51 \tiny$\pm$0.80 &48.79\tiny$\pm$ 1.38 &\textbf{90.91}\tiny$\pm$  2.25  &\textbf{85.67}\tiny$\pm$  6.11 &\textbf{79.65}\tiny$\pm$  0.50   &\textbf{82.65}\tiny$\pm$  0.62 \\
\bottomrule
\end{tabular}
\caption{Performance of baselines trained with augmented train dataset. The scores are averaged from three runs with different seeds.}
\label{tbl:baseline-augment-full}
\end{table*}

\section{Comparison to JointMatch}\label{app:compare-jointmatch}
\begin{figure}[ht]
    \centering
    \begin{subfigure}[b]{.49\textwidth}
    \includegraphics[width=\textwidth]{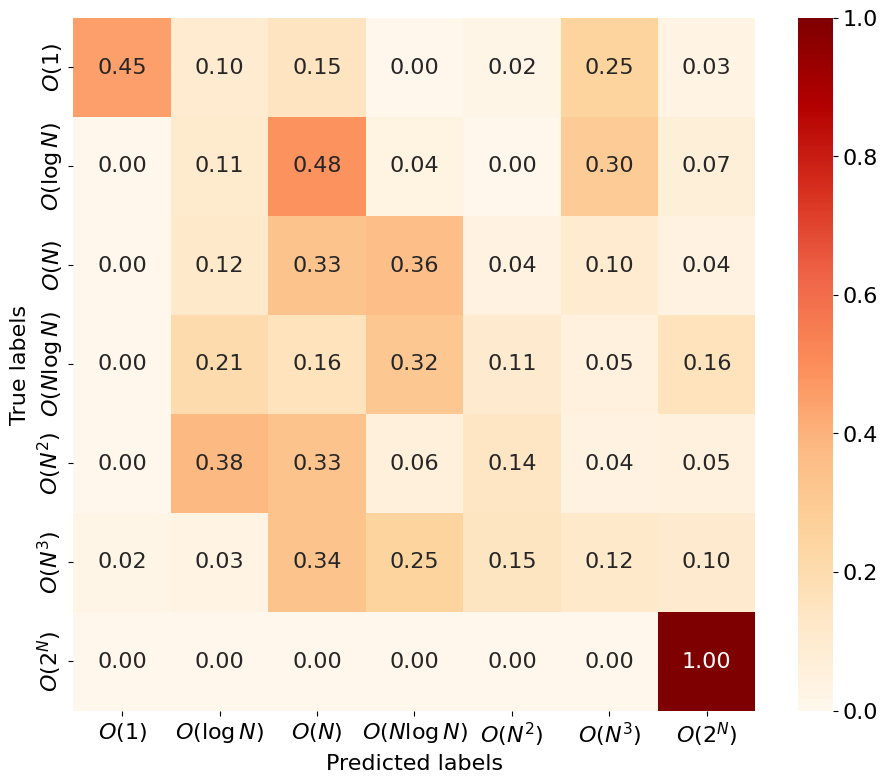}
    \caption{Confusion matrix of CodeT5+ JointMatch }
    \label{fig:codet5p-jointmatch-confusion}
    \end{subfigure}
    \hfill
    \begin{subfigure}[b]{.49\textwidth}
    \includegraphics[width=\textwidth]{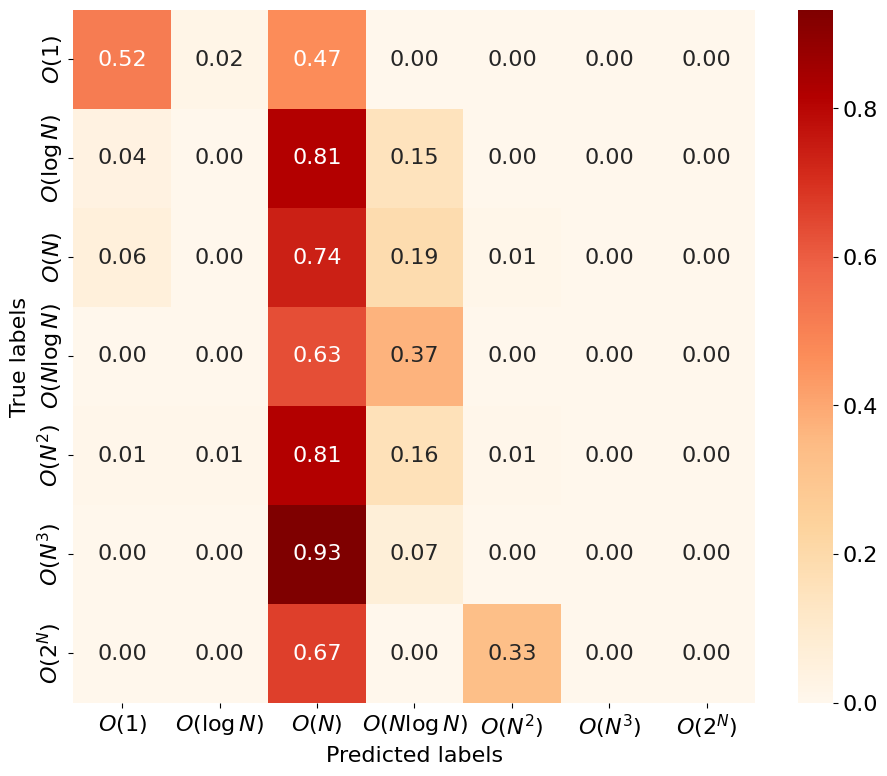}
    \caption{Confusion matrix of UniXcoder JointMatch}
    \label{fig:unixcoder-jointmath-confusion}
    \end{subfigure}
    \caption{CodeT5+, UniXcoder JointMatch performance visualization for CodeComplex (Java) 10-shot.}
\end{figure}

JointMatch is a recognized SSL approach.
It displays the state-of-the-art performance for text classification in low-resource settings.
However, we found this unsuitable for the code time complexity prediction task.
Tables~\ref{tbl:baseline-comparison-acc-full} and~\ref{tbl:baseline-comparison-f1}
illustrates that JointMatch underperforms compared to the standard self-training approaches.
Confusion matrices in Figures~\ref{fig:codet5p-jointmatch-confusion} and~\ref{fig:unixcoder-jointmath-confusion}
provide further details that JointMatch is either biased on several classes or is underfitted.
This is even more remarkable as Figures~\ref{fig:codet5p-confusion} and~\ref{fig:unixcoder-confusion}
displays much better performance.
This poor performance of JointMatch is that the model is primarily developed for text classification datasets
such as AG News, Yahoo! Answers, and IMDB.
These datasets are completely different from the code time complexity prediction datasets
and thus, it is reasonable that the learning mechanism of JointMatch is not effective on the time complexity prediction.

\section{LLM Comparison}\label{app:llm-comparison}
Table~\ref{tbl:llm-comparison} illustrates the accuracy and F1 performance of LLMs and \texttt{TCProF}.
Likewise to the analysis from Section~\ref{ssec:llm-comparison},
\texttt{TCProF} demonstrates performance competitive to the LLMs for F1-scores as well.
\begin{table}[ht]
\centering
\begin{tabular}{lcccccc}
\hline
               &  \multicolumn{2}{c}{CodeComplex~(Java)}  &  \multicolumn{2}{c}{CodeComplex~(Python)}  &        \multicolumn{2}{c}{CorCoD}           \\
               &         Acc.         &         F1          &         Acc.          &         F1           &          Acc.          &         F1          \\
\hline
\hline
GPT3.5              &62.15                &37.96                &   32.55               &     29.24            &    69.42               &      43.29       \\
GPT4                &\textbf{64.01}       &46.90                &   53.04               &     45.15            &   \textbf{78.86}       &      56.95       \\
Gemini-pro          &49.54                &29.73                &   31.05               &     29.76            &    61.91               &      41.26       \\
\hline
\hline
\texttt{TCProF}(UniXcoder)       &53.85  &\textbf{49.45}  &70.29    &53.17   &  63.16  &63.57  \\
\hline
\end{tabular}
\caption{Comparison with the performance of LLM and 10-shot \texttt{TCProF}.}
\label{tbl:llm-comparison}
\end{table}

\section{Usage of $\mathbb{Sym}$ as a Code Time Complexity Classifier}\label{app:symbolic-performance}
Our symbolic module,~$\mathbb{Sym}$, is specifically designed to enhance the pseudo-labeling process
within the \texttt{TCProF}, aiming to improve the code time complexity prediction accuracy.
This module leverages symbolic reasoning to generate more reliable pseudo-labels,
empirically proven in Tables~\ref{tbl:baseline-comparison-acc-part} and~\ref{tbl:ablation-studies}.
For an extensive analysis,
we present the experimental results in Table~\ref{tbl:sym-performance},
using $\mathbb{Sym}$ alone to classify the time complexity of given code snippets.

\begin{table}[ht]
    \centering
    \begin{tabular}{l|cccccc}
    \toprule
                    &  \multicolumn{2}{c}{CodeComplex~(Java)}  &  \multicolumn{2}{c}{CodeComplex~(Python)}  &        \multicolumn{2}{c}{CorCoD}           \\
                    &         Acc.         &         F1        &         Acc.          &         F1         &          Acc.          &         F1          \\
    \hline
    \hline
    $\mathbb{Sym}$  &        49.69         &        40.83      &         55.76         &         41.27      &        53.68           &       37.53   \\
    \bottomrule
    \end{tabular}
    \caption{The performance of $\mathbb{Sym}$ as a time complexity classifier of the code snippets.}
    \label{tbl:sym-performance}
\end{table}

Notably, $\mathbb{Sym}$ shows relatively better performance on Java datasets compared to Python datasets.
This difference can be attributed to the strict syntax of Java, which aids $\mathbb{Sym}$ in more
effective identification and process of loop structures---a critical aspect for time complexity analysis.
Although the overall performance of $\mathbb{Sym}$ is promising,
its integration with the broader \texttt{TCProF},
demonstrates the most effective results.
Appendix~\ref{app:symbolic} illustrates with examples,
the detailed procedure of how $\mathbb{Sym}$ operates.

While $\mathbb{Sym}$ does not capture
every potential operation involved in computing time complexity,
its design is strategically tailored to
assist the baseline model's capability in producing more precise pseudo-labels.
Moving forward, we recognize the potential for further enhancing $\mathbb{Sym}$
by expanding its scope to include a broader range of conditions.
This development is part of our future research
aimed to broaden our understanding of code time complexity prediction.

\section{Symbolic Pseudo-Labeling Running Example}\label{app:symbolic}

\begin{figure}[h!]
    \centering
    \includegraphics[width=\columnwidth]{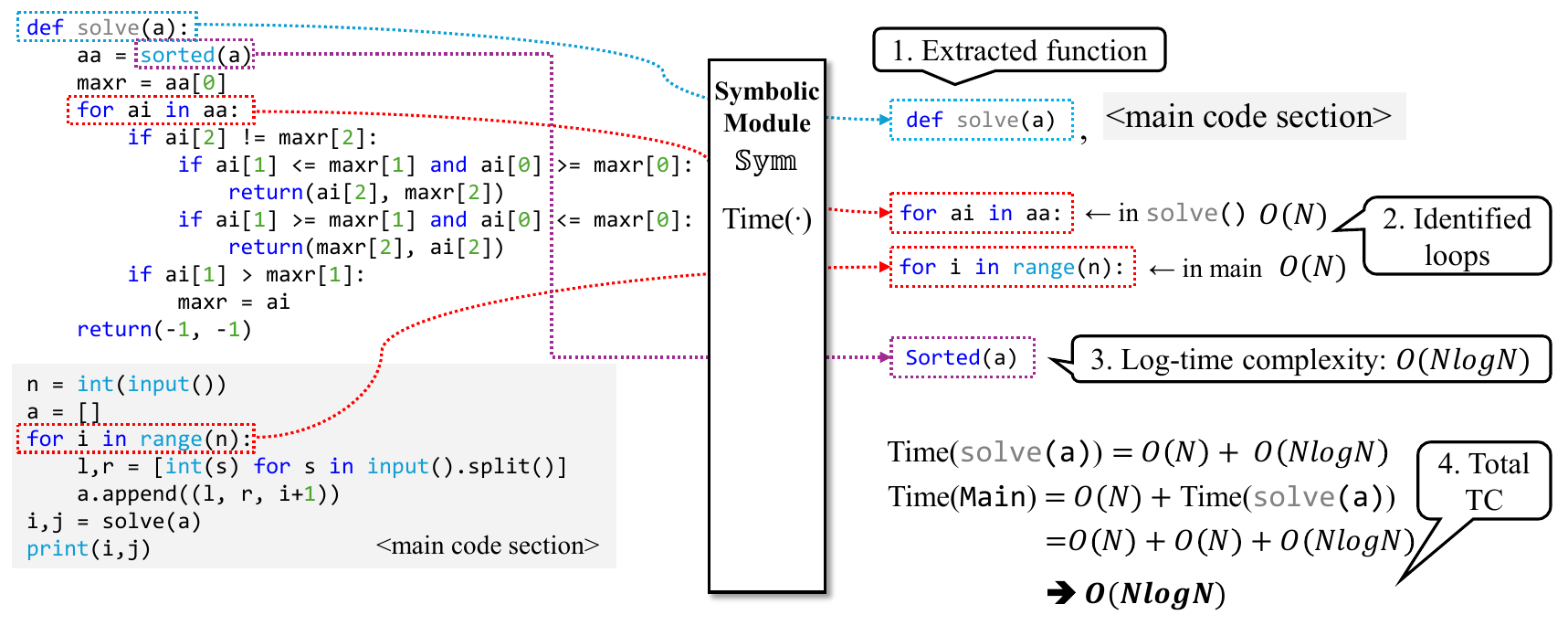}
    \caption{Execution process of the Symbolic Module using Python code snippet.}
    \label{fig:sym_fig}
\end{figure}

In the symbolic module~$\mathbb{Sym}$, the first step involves employing regular expressions~(Regex)
to ascertain the existence of functions within the source code.
Figure~\ref{fig:sym_fig} presents a Python code snippet that 
contains a function named \textcolor{gray}{\texttt{solve}}\texttt{()}.
The module recognizes \textcolor{gray}{\texttt{solve}}\texttt{()} as a function due to 
the presence of the keyword \textcolor{blue}{\texttt{def}}.
Subsequently, $\mathbb{Sym}$ extends its analysis to detect loops and recursion, 
again utilizing Regex.

In the given example, $\mathbb{Sym}$ identifies two for-loops due to 
the the keyword \textcolor{blue}{\texttt{for}}: 
one within the \textcolor{gray}{\texttt{solve}}\texttt{()} function 
and another in the main code section.
In the main code section, $\mathbb{Sym}$ checks \textcolor{gray}{\texttt{solve}}\texttt{()} 
is called only once.
Through Regex matching, the module employs the keyword \textcolor{blue}{\texttt{in}}, 
along with the associated range variables \texttt{aa} and \texttt{n},
to determine the size of each loop.
As depicted in Figure~\ref{fig:sym_fig}, $\mathbb{Sym}$ calculates the time complexity of for loop in 
\textcolor{gray}{\texttt{solve}}\texttt{()} as $O(N)$.
Additionally, the module detects the use of the keyword \textcolor{cyan}{\texttt{sorted}},
which leads to the derivation of the combined time complexity for 
\textcolor{gray}{\texttt{solve}}\texttt{()} as $O(N) + O(N\log N)$.

Finally, $\mathbb{Sym}$ computes the overall time complexity for the main code section.
Given the presence of an additional for loop in this code section,
the module calculates the total time complexity as $O(N) + O(N) + O(N\log N) = O(N\log N)$.

\section{Error Analysis}\label{app:error-analysis}

\begin{figure}[ht]
    \centering
    \begin{subfigure}[b]{.49\textwidth}
    \includegraphics[width=\textwidth]{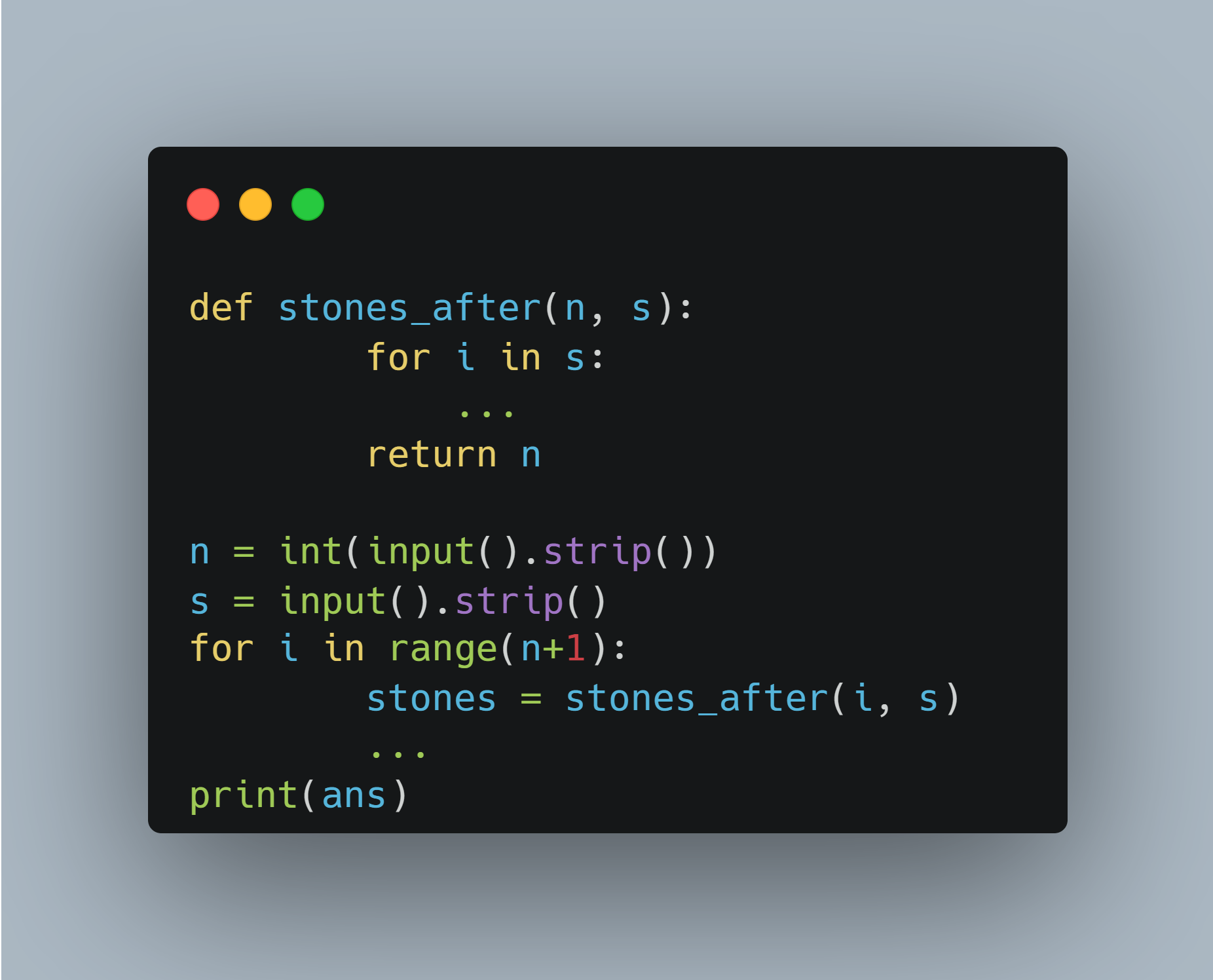}
    \caption{Correct code example predicted $O(N^2)$.}
    \label{fig:correct-code}
    \end{subfigure}
    \hfill
    \begin{subfigure}[b]{.49\textwidth}
    \includegraphics[width=\textwidth]{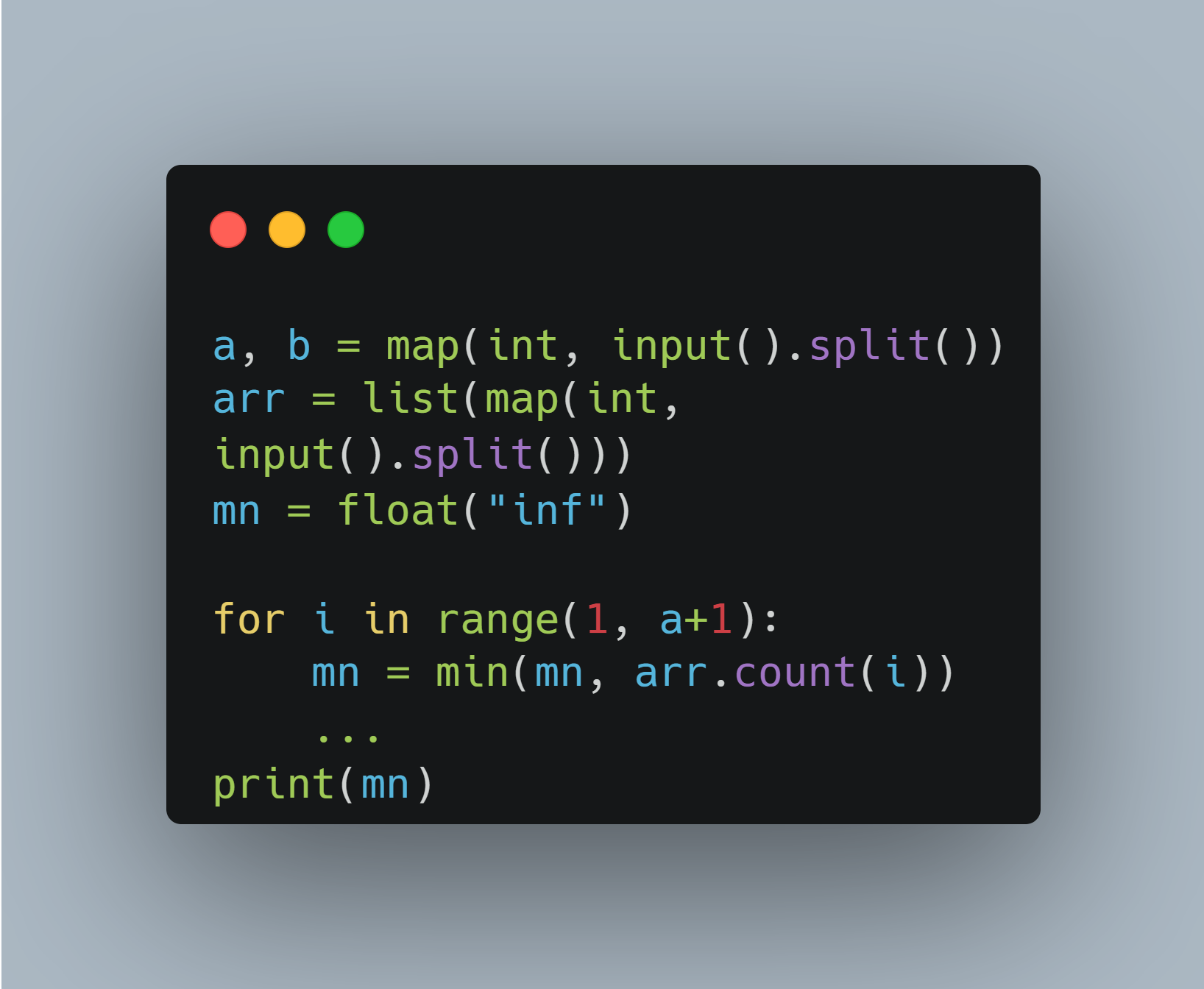}
    \caption{Incorrect code example predicted $O(N)$.}
    \label{fig:incorrect-code}
    \end{subfigure}
    \caption{Code examples of $O(N^2)$~class.}
\end{figure}

From the test cases, we have analyzed the errors, and as discussed in Section~\ref{ssec:error-analysis},
$O(N^2)\rightarrow O(N)$ errors occur the most.
Without sufficient attention to the input size,
it is sometimes confusing, even for human experts, to differentiate two codes, each in $O(N)$ and $O(N^2)$~classes, respectively.
For instance, the code instance in Figure~\ref{fig:correct-code} is rather straightforward.
The \texttt{\textcolor{blue}{stones\_after}} function is called in the main part which runs linear to the input size.
The main part calls the function $n$~times and thus, the time complexity of the whole code in $O(N^2)$.
However, the code in Figure~\ref{fig:incorrect-code} contains \texttt{\textcolor{purple}{count}}
operation which runs linear to the size of \texttt{\textcolor{blue}{arr}}.
If the code does not have the knowledge on the operation,
the model is likely to predict its time complexity as $O(N)$, which is wrong.
This is mostly seen in the codes of class $O(N^2)$ and thus, the majority of errors are
the $O(N^2)\rightarrow O(N)$.

\section{Extensive Analyses on Augmentation}\label{app:aug-analysis}
\begin{table}[h!]
\centering
\resizebox{\textwidth}{!}{
\begin{tabular}{@{}llllllll@{}}
\toprule
\multirow{2}{*}{Models} & \multirow{2}{*}{Augmentation} & \multicolumn{2}{c}{CodeComplex~(Java)}
 & \multicolumn{2}{c}{CodeComplex~(Python)}                                                          
 & \multicolumn{2}{c}{CorCoD}                                                                      \\
& & \multicolumn{1}{c}{5}& \multicolumn{1}{c}{10} & \multicolumn{1}{c}{5}      & \multicolumn{1}{c}{10} & \multicolumn{1}{c}{5} & \multicolumn{1}{c}{10} \\ \midrule
\multirow{5}{*}{CodeBERT}       & $\mathbb{AUG}_{BT}$               & 21.85\tiny$\pm$2.61   & 29.23\tiny$\pm$3.18  & 25.06\tiny$\pm$8.16  & 43.44\tiny$\pm$5.19  & 35.78\tiny$\pm$9.18 & 38.94\tiny$\pm$5.27 \\
                                & $\mathbb{AUG}_{LC\_Natural}$      & 26.32\tiny$\pm$4.20   & 30.20\tiny$\pm$6.21  & 39.82\tiny$\pm$1.56  & 50.61\tiny$\pm$6.05  & 33.33\tiny$\pm$5.40 & 34.03\tiny$\pm$2.43 \\
                                & $\mathbb{AUG}_{LC\_Artificial}$   & 17.52\tiny$\pm$8.20   & 20.06\tiny$\pm$8.13  & 38.86\tiny$\pm$7.60  & 45.42\tiny$\pm$3.64  & 28.42\tiny$\pm$2.11 & 33.68\tiny$\pm$1.05 \\
                                & $\mathbb{AUG}_{BT+LC\_Natural}$   & 25.72\tiny$\pm$0.23   & 38.10\tiny$\pm$1.31  & 39.82\tiny$\pm$8.30  & 55.53\tiny$\pm$10.25 & 36.13\tiny$\pm$1.61 & 41.05\tiny$\pm$1.82 \\
                                & $\mathbb{AUG}_{BT+LC\_Artificial}$& 24.75\tiny$\pm$6.28   & 39.30\tiny$\pm$4.21  & 43.24\tiny$\pm$6.76  & 43.30\tiny$\pm$3.86  & 28.77\tiny$\pm$5.40 & 40.00\tiny$\pm$4.22 \\ 
                                \midrule
\multirow{5}{*}{GraphCodeBERT}  & $\mathbb{AUG}_{BT}$               & 29.00\tiny$\pm$10.99  & 36.91\tiny$\pm$4.62  & 34.22\tiny$\pm$7.28  & 47.40\tiny$\pm$7.63  & 32.65\tiny$\pm$3.82 & 44.56\tiny$\pm$5.30 \\
                                & $\mathbb{AUG}_{LC\_Natural}$      & 25.87\tiny$\pm$3.07   & 24.76\tiny$\pm$1.03  & 29.64\tiny$\pm$2.39  & 47.06\tiny$\pm$7.40  & 36.49\tiny$\pm$2.49 & 45.26\tiny$\pm$8.42 \\
                                & $\mathbb{AUG}_{LC\_Artificial}$   & 22.07\tiny$\pm$1.05   & 30.42\tiny$\pm$2.93  & 33.46\tiny$\pm$15.61 & 46.30\tiny$\pm$6.85  & 34.38\tiny$\pm$3.38 & 41.75\tiny$\pm$8.83 \\
                                & $\mathbb{AUG}_{BT+LC\_Natural}$   & 25.80\tiny$\pm$2.40   & 33.85\tiny$\pm$2.29  & 34.15\tiny$\pm$12.08 & 47.88\tiny$\pm$3.80  & 38.94\tiny$\pm$2.78 & 43.86\tiny$\pm$7.62 \\
                                & $\mathbb{AUG}_{BT+LC\_Artificial}$& 21.10\tiny$\pm$1.79   & 35.46\tiny$\pm$8.07  & 45.21\tiny$\pm$7.89  & 52.59\tiny$\pm$4.69  & 37.89\tiny$\pm$4.59 & 40.65\tiny$\pm$7.40 \\ 
                                \midrule
\multirow{5}{*}{CodeT5+}        & $\mathbb{AUG}_{BT}$               & 20.13\tiny$\pm$3.31   & 26.77\tiny$\pm$5.38  & 39.34\tiny$\pm$19.44 & 45.36\tiny$\pm$11.58 & 34.73\tiny$\pm$7.59 & 38.94\tiny$\pm$2.79 \\
                                & $\mathbb{AUG}_{LC\_Natural}$      & 26.47\tiny$\pm$5.49   & 29.53\tiny$\pm$5.83  & 32.92\tiny$\pm$14.47 & 53.89\tiny$\pm$16.03 & 35.78\tiny$\pm$5.47 & 42.45\tiny$\pm$7.97\\
                                & $\mathbb{AUG}_{LC\_Artificial}$   & 21.25\tiny$\pm$8.79   & 31.09\tiny$\pm$6.45  & 33.33\tiny$\pm$17.78 & 60.86\tiny$\pm$8.22  & 27.37\tiny$\pm$2.78 & 41.40\tiny$\pm$5.80 \\
                                & $\mathbb{AUG}_{BT+LC\_Natural}$   & 26.85\tiny$\pm$1.14   & 35.86\tiny$\pm$4.53  & 33.94\tiny$\pm$3.82  & 57.44\tiny$\pm$1.17  & 36.14\tiny$\pm$4.98 & 43.85\tiny$\pm$6.34 \\
                                & $\mathbb{AUG}_{BT+LC\_Artificial}$& 24.53\tiny$\pm$7.18   & 36.16\tiny$\pm$5.24  & 40.43\tiny$\pm$17.16 & 59.15\tiny$\pm$1.60  & 31.93\tiny$\pm$2.19 & 40.69\tiny$\pm$4.26 \\ 
                                \midrule
\multirow{5}{*}{UniXcoder}      & $\mathbb{AUG}_{BT}$               & 29.75\tiny$\pm$3.17   & 33.40\tiny$\pm$14.43 & 50.31\tiny$\pm$19.27 & 52.93\tiny$\pm$10.66 & 33.15\tiny$\pm$3.72 & 50.17\tiny$\pm$9.43 \\
                                & $\mathbb{AUG}_{LC\_Natural}$      & 30.72\tiny$\pm$8.40   & 33.03\tiny$\pm$6.55  & 44.53\tiny$\pm$13.40 & 53.55\tiny$\pm$4.10  & 38.94\tiny$\pm$3.16 & 47.01\tiny$\pm$4.38 \\
                                & $\mathbb{AUG}_{LC\_Artificial}$   & 22.81\tiny$\pm$6.02   & 42.28\tiny$\pm$16.47 & 40.91\tiny$\pm$5.95  & 50.75\tiny$\pm$8.10  & 37.89\tiny$\pm$8.42 & 43.15\tiny$\pm$3.16 \\
                                & $\mathbb{AUG}_{BT+LC\_Natural}$   & 34.22\tiny$\pm$12.32  & 39.00\tiny$\pm$6.64  & 51.63\tiny$\pm$12.42 & 63.04\tiny$\pm$8.75  & 44.21\tiny$\pm$3.16 & 51.57\tiny$\pm$5.57 \\
                                & $\mathbb{AUG}_{BT+LC\_Artificial}$& 32.73\tiny$\pm$7.54   & 40.19\tiny$\pm$15.17 & 47.33\tiny$\pm$13.84 & 59.15\tiny$\pm$7.17  & 38.59\tiny$\pm$9.78 & 51.93\tiny$\pm$6.77 \\ 
                                \bottomrule
\end{tabular}%
}
\caption{Accuracy of augmentation strategies, $\mathbb{AUG}_{BT}$ and $\mathbb{AUG}_{LC}$ and $\mathbb{AUG}_{BT+LC}$ in natural and artificial settings.
The scores are averaged from three runs with different seeds.}
\label{tbl:extensive-aug-acc}
\end{table}
We have performed an extensive analysis of our augmentation strategy across four models: CodeBERT, GraphCodeBERT, UniXcoder, and CodeT5+. 
As stated in Appendix~\ref{app:baseline-selection}, CodeBERT and GraphcodeBERT are unsuitable for self-training.
Similarly, in this experiment, we notice that CodeBERT and GraphcodeBERT have no consistent tendency of the results. 
In contrast, UniXcoder and CodeT5+ exhibit more consistent performance tendencies.
Therefore, this section focuses on the accuracy and F1-score results of UniXcoder and CodeT5+.

$\mathbb{AUG}_{BT}$ involves augmenting data through Back-Translation of the original labeled code. $\mathbb{AUG}_{LC\_Natural}$ denotes the experimental condition where the initial labeled data is used without specific constraints. In contrast, $\mathbb{AUG}_{LC\_Artificial}$ restricts the initial labeled data to code containing either for loops or while loops, ensuring that all augmented Loop-Conversion data can be generated for every initial labeled data.
The combined strategy $\mathbb{AUG}_{BT+LC\_Natural}$ integrates Back-Translation augmentation with Loop-Conversion augmentation under natural conditions. Similarly, $\mathbb{AUG}_{BT+LC\_Artificial}$ combines Back-Translation augmentation with Loop-Conversion augmentation under artificial conditions.

\begin{table}[h!]
\centering
\resizebox{\textwidth}{!}{
\begin{tabular}{@{}llllllll@{}}
\toprule
\multirow{2}{*}{Models} & \multirow{2}{*}{Augmentation} & \multicolumn{2}{c}{CodeComplex~(Java)}
 & \multicolumn{2}{c}{CodeComplex~(Python)}                                                          
 & \multicolumn{2}{c}{CorCoD}                                                                      \\
& & \multicolumn{1}{c}{5}& \multicolumn{1}{c}{10} & \multicolumn{1}{c}{5}      & \multicolumn{1}{c}{10} & \multicolumn{1}{c}{5} & \multicolumn{1}{c}{10} \\ \midrule
\multirow{5}{*}{CodeBERT}      & $\mathbb{AUG}_{BT}$                & 19.94\tiny$\pm$5.56  & 27.48\tiny$\pm$6.51  & 19.90\tiny$\pm$5.09 & 25.46\tiny$\pm$7.78 & 33.45\tiny$\pm$8.85  & 35.44\tiny$\pm$10.80 \\
                               & $\mathbb{AUG}_{LC\_Natural}$       & 28.54\tiny$\pm$0.60  & 34.38\tiny$\pm$1.04  & 22.75\tiny$\pm$5.23 & 33.69\tiny$\pm$5.40 & 33.09\tiny$\pm$10.91 & 24.47\tiny$\pm$3.93  \\
                               & $\mathbb{AUG}_{LC\_Artificial}$    & 15.22\tiny$\pm$7.17  & 15.78\tiny$\pm$5.74  & 27.81\tiny$\pm$8.68 & 33.00\tiny$\pm$6.67 & 18.37\tiny$\pm$3.60  & 32.31\tiny$\pm$7.57  \\
                               & $\mathbb{AUG}_{BT+LC\_Natural}$    & 23.95\tiny$\pm$1.42  & 28.23\tiny$\pm$0.36  & 28.34\tiny$\pm$3.43 & 35.95\tiny$\pm$7.99 & 29.80\tiny$\pm$5.94  & 46.39\tiny$\pm$3.01 \\
                               & $\mathbb{AUG}_{BT+LC\_Artificial}$ & 22.45\tiny$\pm$7.29  & 34.14\tiny$\pm$6.96  & 28.19\tiny$\pm$7.91 & 34.12\tiny$\pm$6.02 & 22.30\tiny$\pm$1.45  & 41.05\tiny$\pm$4.25 \\
                               \midrule
\multirow{5}{*}{GraphCodeBERT} & $\mathbb{AUG}_{BT}$                & 24.77\tiny$\pm$12.26 & 29.14\tiny$\pm$5.32  & 25.78\tiny$\pm$4.32 & 33.16\tiny$\pm$2.92 & 27.36\tiny$\pm$6.60  & 44.89\tiny$\pm$5.06  \\
                               & $\mathbb{AUG}_{LC\_Natural}$       & 26.43\tiny$\pm$6.25  & 22.30\tiny$\pm$2.24  & 25.75\tiny$\pm$9.66 & 33.39\tiny$\pm$1.79 & 34.29\tiny$\pm$8.78  & 43.88\tiny$\pm$4.57  \\
                               & $\mathbb{AUG}_{LC\_Artificial}$    & 15.59\tiny$\pm$6.34  & 22.18\tiny$\pm$3.79  & 26.07\tiny$\pm$5.80 & 36.00\tiny$\pm$3.21 & 40.69\tiny$\pm$4.65  & 41.71\tiny$\pm$7.23  \\
                               & $\mathbb{AUG}_{BT+LC\_Natural}$    & 20.69\tiny$\pm$4.23  & 35.64\tiny$\pm$7.23  & 27.33\tiny$\pm$5.72 & 35.99\tiny$\pm$4.39 & 38.17\tiny$\pm$1.78  & 40.76\tiny$\pm$6.64 \\
                               & $\mathbb{AUG}_{BT+LC\_Artificial}$ & 20.98\tiny$\pm$1.78  & 34.34\tiny$\pm$2.00  & 31.30\tiny$\pm$3.69 & 42.97\tiny$\pm$2.52 & 40.96\tiny$\pm$4.84  & 36.77\tiny$\pm$5.18 \\
                               \midrule
\multirow{5}{*}{CodeT5+}       & $\mathbb{AUG}_{BT}$                & 22.90\tiny$\pm$1.60  & 27.31\tiny$\pm$3.13  & 31.94\tiny$\pm$9.17 & 35.83\tiny$\pm$4.52 & 35.80\tiny$\pm$10.98 & 43.85\tiny$\pm$2.99  \\
                               & $\mathbb{AUG}_{LC\_Natural}$       & 25.33\tiny$\pm$3.56  & 26.42\tiny$\pm$4.02  & 26.60\tiny$\pm$5.77 & 42.50\tiny$\pm$8.51 & 39.36\tiny$\pm$6.15  & 46.76\tiny$\pm$6.72  \\
                               & $\mathbb{AUG}_{LC\_Artificial}$    & 16.67\tiny$\pm$4.92  & 27.68\tiny$\pm$8.33  & 34.73\tiny$\pm$5.42 & 43.37\tiny$\pm$4.71 & 27.92\tiny$\pm$3.56  & 44.87\tiny$\pm$7.29 \\
                               & $\mathbb{AUG}_{BT+LC\_Natural}$    & 26.23\tiny$\pm$2.07  & 30.69\tiny$\pm$ 8.08 & 31.12\tiny$\pm$5.61 & 42.38\tiny$\pm$1.11 & 41.78\tiny$\pm$5.12  & 47.98\tiny$\pm$4.70 \\
                               & $\mathbb{AUG}_{BT+LC\_Artificial}$ & 21.70\tiny$\pm$5.73  & 32.92\tiny$\pm$3.43  & 33.27\tiny$\pm$2.25 & 43.07\tiny$\pm$1.69 & 30.77\tiny$\pm$8.12  & 48.19\tiny$\pm$4.78 \\ 
                               \midrule
\multirow{5}{*}{UniXcoder}     & $\mathbb{AUG}_{BT}$                & 29.18\tiny$\pm$3.95  & 33.52\tiny$\pm$10.03 & 31.43\tiny$\pm$5.98 & 39.24\tiny$\pm$3.60 & 38.82\tiny$\pm$9.97  & 52.39\tiny$\pm$8.03  \\
                               & $\mathbb{AUG}_{LC\_Natural}$       & 27.24\tiny$\pm$2.77  & 36.26\tiny$\pm$16.28 & 29.48\tiny$\pm$3.87 & 34.82\tiny$\pm$4.26 & 40.33\tiny$\pm$8.51  & 48.20\tiny$\pm$13.78 \\
                               & $\mathbb{AUG}_{LC\_Artificial}$    & 28.08\tiny$\pm$13.97 & 35.58\tiny$\pm$8.97  & 31.11\tiny$\pm$7.11 & 35.81\tiny$\pm$4.66 & 38.41\tiny$\pm$9.57  & 44.36\tiny$\pm$7.43  \\
                               & $\mathbb{AUG}_{BT+LC\_Natural}$    & 31.02\tiny$\pm$14.22 & 34.21\tiny$\pm$10.92 & 33.53\tiny$\pm$5.73 & 45.95\tiny$\pm$8.69 & 43.42\tiny$\pm$1.29  & 52.57\tiny$\pm$5.91 \\
                               & $\mathbb{AUG}_{BT+LC\_Artificial}$ & 29.34\tiny$\pm$6.21  & 38.99\tiny$\pm$5.56  & 30.94\tiny$\pm$6.00 & 42.07\tiny$\pm$4.98 & 40.62\tiny$\pm$10.50 & 51.83\tiny$\pm$5.38 \\                               
                               \bottomrule
\end{tabular}%
}
\caption{F1-scores of augmentation strategies, $\mathbb{AUG}_{BT}$ and $\mathbb{AUG}_{LC}$ and $\mathbb{AUG}_{BT+LC}$ in natural and artificial settings.
The scores are averaged from three runs with different seeds.}
\label{tbl:extensive-aug-f1}
\end{table}

We analyzed the results in two parts: $\mathbb{AUG}_{LC\_Natural}$ vs. $\mathbb{AUG}_{LC\_Artificial}$ and $\mathbb{AUG}_{BT}$  vs. $\mathbb{AUG}_{LC\_Natural}$.

The first comparison, $\mathbb{AUG}_{LC\_Natural}$ and $\mathbb{AUG}_{LC\_Artificial}$, highlights two different augmentation strategies within $\mathbb{AUG}_{LC\_Natural}$.
Notably, $\mathbb{AUG}_{LC\_Natural}$ significantly enhances model performance compared to $\mathbb{AUG}_{LC\_Artificial}$, particularly in the UniXcoder model.
The average number of data points used for training in the $\mathbb{AUG}_{LC\_Natural}$ is 65, while the $\mathbb{AUG}_{LC\_Artificial}$ uses 70 data in 5-shot settings on the CodeComplex dataset, which has seven labels.
Despite having less training data, $\mathbb{AUG}_{LC\_Natural}$ achieves higher performance.

The second comparison, $\mathbb{AUG}_{BT}$ vs. $\mathbb{AUG}_{LC\_Natural}$, evaluates the differences between $\mathbb{AUG}_{BT}$ and $\mathbb{AUG}_{LC\_Natural}$.
In Tables~\ref{tbl:extensive-aug-acc} and~\ref{tbl:extensive-aug-f1},
$\mathbb{AUG}_{LC\_Natural}$ consistently outperforms $\mathbb{AUG}_{BT}$ in both accuracy
and F1-score across all evaluated models while maintaining lower standard deviations, indicating more reliable and stable performance.
The average number of data points used for training in the $\mathbb{AUG}_{LC\_Natural}$ is 65, while the $\mathbb{AUG}_{BT}$ uses 70 data in 5-shot settings on the CodeComplex dataset, which has seven labels.
Despite having less training data, $\mathbb{AUG}_{LC\_Natural}$ achieves higher performance.

For the CodeT5+ model, $\mathbb{AUG}_{LC\_Natural}$ achieves higher accuracy except in the CodeComplex Python 5-shot setting, where a high standard deviation of 19.44\% was observed in $\mathbb{AUG}_{BT}$.
Similarly, for the UniXcoder model, $\mathbb{AUG}_{LC\_Natural}$ achieves higher accuracy in most cases, except in the CodeComplex Python 5-shot setting, where it shows 6\%p lower accuracy. 
These large performance differences are likely due to the inherent variability in the CodeComplex Python data results, contributing to the instability observed across all models. 
Despite this 6\%p lower standard deviation compared to $\mathbb{AUG}_{BT}$ in the CodeComplex Python 5-shot setting, indicating more stable performance even with a slight drop in accuracy.
In the 10-shot setting, where the data is more comprehensive, $\mathbb{AUG}_{LC\_Natural}$ outperforms $\mathbb{AUG}_{BT}$ in accuracy.

There are some exceptions, such as in the CodeComplex Java 10-shot setting, likely due to the limited diversity of training data.
$\mathbb{AUG}_{LC\_Natural}$ focuses on converting loop parts rather than making substantial code changes, while $\mathbb{AUG}_{BT}$ enhances the diversity of the training data.
When $\mathbb{AUG}_{BT}$ is added to the training data, it helps bridge the performance gap observed in these exceptional cases, as evidenced by the results for $\mathbb{AUG}_{BT+LC\_Artificial}$ in  CodeComplex Java 10-shot setting.
These findings suggest that $\mathbb{AUG}_{LC\_Natural}$ is the most effective augmentation strategy for enhancing the performance of code models.
Moreover, combining $\mathbb{AUG}_{BT}$ with $\mathbb{AUG}_{LC\_Natural}$ further boosts the performance of the models.

\section{Augmentation Prompts}\label{app:aug-prompt}
The augmentation prompts presented in this section are designed to enhance the capabilities
of language models in code Back-Translation and Loop-Conversion tasks.
The first prompt, Back-Translation, challenges the model to translate Java code into Python and then back into Java,
ensuring functional equivalence without syntactical similarity to the original code.
The second prompt, Loop-Conversion, requires the model to convert all `for' loops to `while' loops and vice versa in the given Java code, even at the cost of readability.
To illustrate the effectiveness of these methods, we provide both original code and augmented code examples for each augmentation technique.
Note that we only show parts of the codes because they are too long to display.

\subsection{Back-Translation Examples}
The following is an original example of a Java code.

\begin{javacode}
import ...
public class ProblemD {
	public static void main(String[] args) throws IOException {
		BufferedReader s = new BufferedReader(new InputStreamReader(System.in));
		PrintWriter out = new PrintWriter(System.out);
		String[] data = s.readLine().split(" ");
		String[] line = s.readLine().split(" ");
            double ans = solve(a, v, l, d, w);
		out.println(String.format("\%.07f", ans));
		out.flush();
	}
	private static double solve(double a, double v, double l, double d, double w) {
		double maxSpeedAtD = Math.sqrt(2 * d / a) * a;
		if (v <= w || maxSpeedAtD <= w) {
			// okay. just go
                ...
		}
		double time = 0.0d;
		double maxSpeedTime = Math.sqrt((d / a) + (w * w / (2 * a * a)));
		double maxSpeed = maxSpeedTime * a;
		if (maxSpeed <= v) {
                ...
		}
		
		// after that, just go.
		double timeToV = (v - w) / a;
		double timeToVLen = timeToV * w + 0.5 * timeToV * (v - w);
		if (timeToVLen <= l - d) {
                ...
		}
		return time;
	}
}
\end{javacode}

The following is a transformed example of a Java code.
\begin{javacode}
import ...

public class ProblemD {
	public static void main(String[] args) throws IOException {
		BufferedReader s = new BufferedReader(new InputStreamReader(System.in));
		PrintWriter out = new PrintWriter(System.out);

		String[] data = s.readLine().split(" ");
		String[] line = s.readLine().split(" ");

		double ans = solve(a, v, l, d, w);
		out.println(String.format("%.07f", ans));
		
		out.flush();
	}

	private static double solve(double a, double v, double l, double d, double w) {
		double maxSpeedAtD = Math.sqrt(2 * d / a) * a;
		if (v >= w && maxSpeedAtD > w) {
                ...
		}
	}

	private static double findTimeToReachL(double a, double v, double l) {
            ...
		return timeToMaxSpeed + leftDist / v;
	}

	private static double calculateTimeToReachDestination(double a, double v, double l, double d, double w) {
            ...
		return maxSpeed <= v ? maxSpeedTime + (a * maxSpeedTime - w) / a : calculateTimeForCaseThree(a, v, d, w, maxSpeedTime);
	}

	private static double calculateTimeForCaseThree(double a, double v, double d, double w, double maxSpeedTime) {
            ...
		return v > w ? v / a + time : v / a + vtime + (v - w) / a;
	}
}
\end{javacode}


\subsection{Loop-Conversion Examples}

The following is an original example of a Java code.
\begin{javacode}
import java.util.*;
import java.io.*;

public class Soldiers {
	public static void main(String[] args) throws IOException {
		new Soldiers().run();
	}

	void run() throws IOException {
		br = new BufferedReader(new InputStreamReader(System.in));
		pw = new PrintWriter(System.out);

		int n = nextInt();
		pw.println(3 * (n / 2));

		pw.close();
	}

	BufferedReader br;
	StringTokenizer st;
	PrintWriter pw;

	String next() throws IOException {
		while (st == null || !st.hasMoreTokens()) {
			st = new StringTokenizer(br.readLine());
		}
		return st.nextToken();
	}

	int nextInt() throws IOException {
		return Integer.parseInt(next());
	}

	double nextDouble() throws IOException {
		return Double.parseDouble(next());
	}
}

\end{javacode}

The following is a transformed example of a Java code.
\begin{javacode}
import java.util.*;
import java.io.*;

public class Soldiers {
	public static void main(String[] args) throws IOException {
		new Soldiers().run();
	}

	void run() throws IOException {
		br = new BufferedReader(new InputStreamReader(System.in));
		pw = new PrintWriter(System.out);

		int n = nextInt();
		pw.println(3 * (n / 2));

		pw.close();
	}

	BufferedReader br;
	StringTokenizer st;
	PrintWriter pw;

	String next() throws IOException {
		for(; st == null || !st.hasMoreTokens(); st = new StringTokenizer(br.readLine())) {
		}
		return st.nextToken();
	}

	int nextInt() throws IOException {
		return Integer.parseInt(next());
	}

	double nextDouble() throws IOException {
		return Double.parseDouble(next());
	}
}
\end{javacode}

\newpage

\subsection{Back-Translation Prompt}

\begin{figure}[ht!]
\begin{tcolorbox}[colback=blue!5,colframe=blue!40!black,title=Back-Translation Prompt,left=0.5mm,right=0.5mm,bottom=0.5mm,top=0.5mm,enlarge bottom by=-0.2cm]
You are a powerful AI that can translate Java code into Python code and then back into Java code. \\
I will provide you with a piece of Java code. \\
Your task is to generate an equivalent Python code and then translate it back into Java. \\
When translating back into Java, the code should not be identical to the original Java code, but it should still be functionally equivalent. \\
Please ensure that the output is valid Java and does not contain any syntax or constructs from the other language.
The Java output should be in the following JSON format:\\
\{ \\
    \indent "back-translation": "[Your transformed Java code here]" \\
\}. \\
Please note that "[Your transformed Java code here]" should be replaced with your transformed Java code. \\
Do not include any other keys in the "back-translation" value.\\
Please ensure that you generate the complete transformed Java code, do not stop halfway through.\\
Do not generate Python code, only Java code.\\

Given Java Code:
\begin{javacode}
for (int i = 0; i < 5; i++) { 
 System.out.println("Number is " + i);
}
\end{javacode}

Your Transformed Python Code(Java to Python):

\begin{pythoncode}
for i in range(5):
    print("Number is " + str(I)) 
\end{pythoncode}

Your Transformed Java Code(Python to Java):

\begin{javacode}
int i = 0; 
while (i < 5) { 
    System.out.println("Number is " + i);
    i++; 
} 
\end{javacode}

Here is the Java code:"[Original Java Code]" \\
Please translate this Java code into Python code and then back into Java code, and generate your answer in the following JSON format:\\
\{\{ \\
    "back-translation": "[The transformed Java code]" \\
\}\} \\

Don't cut me off in the middle, create it all the way through.
The output should not contain any text, only code. This includes avoiding explanations, comments, or any other form of text.

Please ensure that the transformed code is properly indented for readability.

Do not generate Python code, only Java code.

\end{tcolorbox}
\caption{LLM prompt examples used in Back-translation.}
\end{figure}

\newpage
\subsection{Loop-Conversion Prompt}

\begin{figure}[h!]
\begin{tcolorbox}[colback=blue!5,colframe=blue!40!black,title=Loop-Conversion Prompt,left=0.5mm,right=0.5mm,bottom=0.5mm,top=0.5mm,enlarge bottom by=-0.2cm, enhanced]
...\\
Your task is to convert all `for' loops into `while' loops and all `while' loops into `for' loops. \\
For example, if the Java code contains a `for' loop, you should change it into a `while' loop. \\
If the Java code contains a `while' loop, you should change it into a `for' loop. \\
If the Java code contains both `for' and `while' loops, you should change all `for' loops into `while' loops and all `while' loops into `for' loops. \\
Please note that all `while' loops should be converted into `for' loops, even if it breaks the readability of the code.
Please note that all `while' loops should be converted into `for' loops, even if it only runs once.
This may result in `for' loops with empty initialization or increment sections. \\
The output should be valid Java code and should not contain any syntax or constructs from other languages. \\
The Java output should be in the following JSON format:\\
Please ensure that your transformed Java code is enclosed in double quotes and ends with a closing quote. \\
\{ \\
    "forwhile": "[Your transformed Java code here]" \\
\}.\\
...\\
Here is an example of how to convert a `for' loop to a `while' loop in Java:\\
...\\
And here is an example of how to convert a Java code that contains both `for' and `while' loops:
\begin{javacode}
Given Java Code:

for(int i = 0; i < 10; i++) {
    System.out.println("For loop: " + i);
}
while(!st.hasMoreTokens()) {
    st = new StringTokenizer(in.readLine());
}  
\end{javacode}
Your Transformed Java Code:  \\
\begin{javacode}
int i = 0;
while(i < 10) {
    System.out.println("For loop: " + i);
    i++;
}
for(; !st.hasMoreTokens(); st = new StringTokenizer(in.readLine())) {
}  
\end{javacode}

Here is the Java code: "[Original Java Code]".\\
Please convert all `for' loops into `while' loops and all `while' loops into `for' loops, and generate your answer in the following JSON format:\\
\{\{\\
    "forwhile": "[The transformed Java code]"\\
\}\}.\\
Don't cut me off ...

\end{tcolorbox}
\caption{LLM prompt examples used in Loop-conversion.}
\label{fig:prompt}
\end{figure}

\newpage
\section{LLM 5-shot In-Context Learning Prompts}\label{app:llm-prompt}
\begin{figure}[ht!]
\begin{tcolorbox}[colback=blue!5,colframe=blue!40!black,title=LLM Prompt,left=0.5mm,right=0.5mm,bottom=0.5mm,top=0.5mm,enlarge bottom by=-0.2cm]
You are the best programmar in the world.
You will be asked to determine the time complexity of the following code.
For the time complexity, choose one time complexity from the following options 'constant', 'logn', 'linear', 'nlogn', 'quadratic', 'cubic', and 'exponential'.
Do not hesitate to use any other supplementary materials you need for the task.
I will first give you the code.
After you read the code,
I will ask you to compute the time complexity of the code.
The following are the demonstrations of the time complexity for codes:\\
----------------------------------------\\
\begin{pythoncode}
n = int(input())
for i in range(n): print(i)
\end{pythoncode}

"complexity": linear\\
----------------------------------------\\
\begin{pythoncode}
print(int(input()))
\end{pythoncode}

"complexity": constant\\
----------------------------------------\\
\begin{pythoncode}
print("*")
print("**")
print("***")
\end{pythoncode}

"complexity": constant\\
----------------------------------------\\
\begin{pythoncode}
n = int(input())
items = list(map(int, input().split()))
items.sort()
\end{pythoncode}

"complexity": nlogn\\
----------------------------------------\\
\begin{pythoncode}
def powerset(items):
    n = len(items)
    for i in range(1 << n):
        subset = []
        for j in range(n):
            if i \& (1 << j):
                subset.append(items[j])
        print(subset)
items = list(map(int, input().split()))
powerset(items)
\end{pythoncode}

"complexity": exponential\\
----------------------------------------\\

Please output the time complexity of the whole code in a json format.
Json format should be\\
\{ \\
    "complexity": time complexity of the whole code\\
\}.\\

\end{tcolorbox}
\caption{LLM prompt used for 5-shot in-context learning evaluation.}
\end{figure}

\end{document}